\documentclass[aps,prl,reprint,groupedaddress,showpacs,amsmath,amssymb,twocolumn,floatfix]{revtex4-1}
\usepackage{graphicx}
\usepackage{bm}
\usepackage{color}
\usepackage[normalem]{ulem}

\usepackage{amsmath}
\usepackage{hyperref}
\usepackage{cleveref}
\usepackage{cancel}
\usepackage{url}

\newcommand{\dnu}{$\Delta \nu$}
\newcommand{\nutot}{$\nu_{\text{total}}$}

\newcommand{\CFn}{$_{0}^{2}$CF }

\newcommand{\CFi}{$_{1}^{2}$CF }

\newcommand{\CFii}{$_{2}^{2}$CF }

\begin{document}

\title{Excitons in the Fractional Quantum Hall Effect}

\author{Naiyuan J. Zhang$^{1}$}
\thanks{These authors contributed equally to this work.}
\author{Ron Q. Nguyen$^{1}$}
\thanks{These authors contributed equally to this work.}
\author{Navketan Batra$^{1,2}$}
\thanks{These authors contributed equally to this work.}
\author{Xiaoxue Liu$^{1}$}
\thanks{Current affliation: Tsung-Dao Lee Institute and School of Physics and Astronomy, Shanghai Jiao Tong University, Shanghai 200240, China}
\author{Kenji Watanabe$^{3}$}
\author{Takashi Taniguchi$^{4}$}
\author{D. E. Feldman$^{1,2}$}
\author{J.I.A. Li$^{1}$}
\thanks{Email: jia$\_$li@brown.edu}

\affiliation{$^{1}$Department of Physics, Brown University, Providence, Rhode Island 02912, USA}
\affiliation{$^2$Brown Theoretical Physics Center, Brown University, Providence, Rhode Island 02912, USA}
\affiliation{$^{3}$Research Center for Electronic and Optical Materials, National Institute for Materials Science, 1-1 Namiki, Tsukuba 305-0044, Japan}
\affiliation{$^{4}$Research Center for Materials Nanoarchitectonics, National Institute for Materials Science,  1-1 Namiki, Tsukuba 305-0044, Japan}

\date{\today}

\maketitle

\textbf{Excitons, Coulomb-driven bound states of electrons and holes, are typically composed of integer charges. However, in bilayer systems influenced by charge fractionalization, a more exotic form of interlayer exciton can emerge, where pairing occurs between constituents that carry fractional charges. Despite numerous theoretical predictions for such fractional excitons, their experimental observation has remained elusive. Here, we report transport signatures of excitonic pairing within fractional quantum Hall effect states. By probing the composition of these excitons and their impact on the underlying wavefunction, we uncover two novel quantum phases of matter. One of these orders can be viewed as the fractional counterpart of the exciton condensate at a total filling of one, while the other involves a more unusual type of exciton that obeys fermionic and anyonic quantum statistics, challenging the standard paradigm of bosonic excitons.}

\begin{figure*}
\includegraphics[width=1\linewidth]{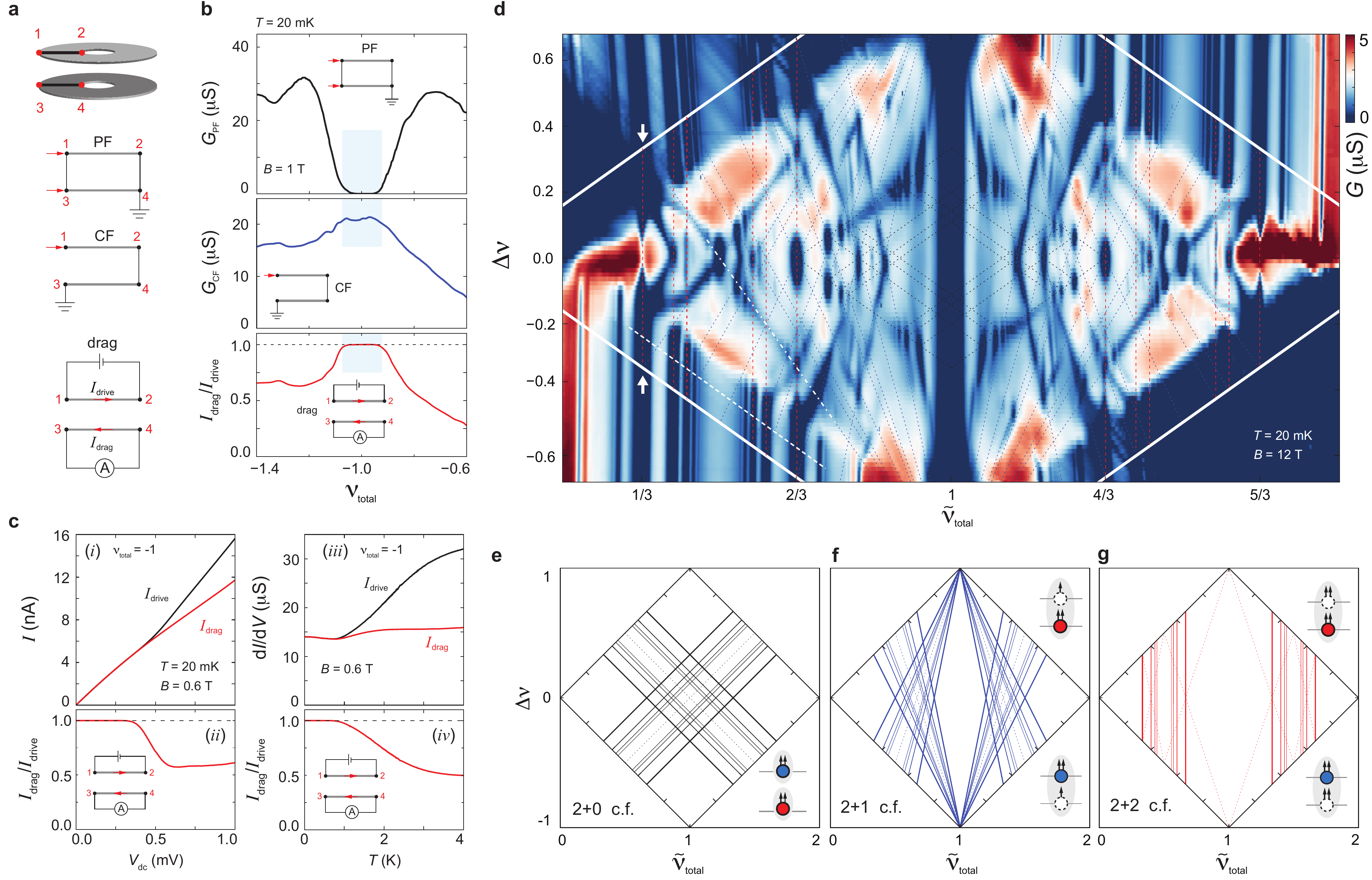}
\caption{\label{fig1}{\bf{Excitonic pairing and fractional quantum Hall effect in quantum Hall bilayer.}} (a) Schematic of the edgeless Corbino geometry used for transport measurements in parallel flow (PF), counterflow, and drag configurations. (b) Transport response measured at \nutot $=-1$, at $B=1$ T and $T = 20$ mK, in the PF geometry $G_{\text{PF}}$ (top panel), counterflow geometry $G_{\text{CF}}$ (middle panel), and the drag ratio $I_{\text{drag}}/I_{\text{drive}}$ (bottom panel). (c) Drag response at $\nu_{\text{total}} = -1$ as a function of d.c. voltage bias $V_{\text{dc}}$ (left) and temperature $T$ (right). (d) $G_{\text{PF}}$ mapped as a function of $\tilde{\nu}_{\text{total}}$ and $\Delta\nu$ at $B=12$ T and $T = 20$ mK. White dashed and dash-dotted lines highlight two FQHE sequences to be examined in detail. (e-g) Schematic diagrams showing FQHE associated with the (e) \CFn, (f) \CFi, and (g) \CFii\ constructions. }
\end{figure*}

Electric charges are typically multiples of a universal unit, the charge of an electron. However, in the fractional quantum Hall effect (FQHE) observed in two-dimensional electronic systems, quasiparticles carry only a fraction of this fundamental charge due to the interplay between correlation and topology ~\cite{review-FH}. These fractionally charged quasiparticles follow anyonic quantum statistics, setting them apart from conventional fermions and bosons. 
The process of charge fractionalization becomes more complex when electrons gain additional quantum numbers such as spin, valley isospin, and layer pseudo-spin. The presence of multiple electron species leads to a multi-component wavefunction, first described by the Halperin two-component wavefunction ~\cite{Halperin1983}, and gives rise to  fractionally charged quasiparticles with an internal degrees of freedom ~\cite{Yang.94,Moon1995,Girvin1996multi}.

In the presence of multiple electron species, the formation of excitons adds a rich new dimension to the FQHE landscape ~\cite{Laughlin1984exciton,Wen1992,Kamilla1997exciton,Yang1994quasiexciton,Park2000fractionalexciton,Park2001fractionalexciton,Jain2005fractionalexciton,Quinn2006quasiexciton,Barkeshi2018,Zhang2023fractionalexciton,Kwan2022quasiexciton,faugno2020}. 
Previous experimental studies have demonstrated the formation of interlayer excitons by confining charge carriers in closely separated but electrically isolated graphene layers, particularly at a total integer Landau level filling in quantum Hall bilayer structures. These interlayer excitons are composed of constituents with integer charges, and due to their bosonic nature, they condense into a Bose-Einstein condensate at low temperatures. This state can be detected through counterflow drag experiments, where a perfect drag response indicates the presence of the condensate ~\cite{Eisenstein2014,Li2017superfluid,Liu2017superfluid,Liu2022crossover}.

Moving away from the total integer filling factor, the coexistence of excitons and FQHE has been the focus of extensive theoretical discussions. An example of this is the $(nnn)$ wavefunction,  which represents a fractional analogue of exciton condensates in quantum Hall bilayers ~\cite{Wen1992}. Additionally, at fractional filling fractions, excitons  could form between fractionally charged constituents, leading to the emergence of excitons that follow non-bosonic statistics and behave like fermions or anyons ~\cite{Barkeshi2018}. This could result in unique ground states that are fundamentally different from the traditional Bose-Einstein condensation.

Despite extensive theoretical discussions ~\cite{Laughlin1984exciton,Wen1992,Kamilla1997exciton,Yang1994quasiexciton,Park2000fractionalexciton,Park2001fractionalexciton,Jain2005fractionalexciton,Quinn2006quasiexciton,Barkeshi2018,Zhang2023fractionalexciton,Kwan2022quasiexciton,faugno2020}, coexistence of excitons and the FQHE has, to the best of our knowledge, remained unexplored experimentally. This leaves the nature of fractional excitons as open questions. These unresolved issues motivate this current report, which is structured as follows: we begin by testing perfect drag at a total integer filling factor. In agreement with previous studies, we observe transport behavior indicative of an exciton condensate known as the (111) state. This condensate persists even with significant charge imbalances between the two layers, as long as the overall filling factor remains constant, which is represented in Fig.~\ref{fig1}. Next, we identify two classes of FQHE states away from the integer filling, both of which exhibit perfect drag response that is indicative of excitonic pairing. The first class extends along lines of constant fractional overall filling factors over a range of layer imbalance. Similar to the (111) state, these states can be explained by fractional analogues of exciton condensates.  The second class of FQHE is stable only around specific points defined by the filling factor of each layer, corresponding to a bilayer generalization of Jain's composite fermion states. We attribute this behavior to the presence of excitonic quasiparticles, which, at certain filling factors, exhibit fractional statistics. 

The results reported in the main text are further substantiated by detailed discussions in the Methods section. This section offers a comprehensive review, summarizing the notations employed and elaborating on the measurement configurations as well as the inter-layer degrees of freedom involved in the experiments. The theoretical methods section explores the relevant wave functions, the conditions required for perfect drag, possible quantum Hall states on the observed plateaus, the constituent charges of excitons, and their quantum statistics. For additional information and in-depth analysis, readers are directed to the Supplementary Materials ~\cite{SI}.

First, we will discuss the transport  signatures of the exciton condensate at an integer total filling of quantum Hall graphene bilayer.
To eliminate the influence of edge channels, the samples are shaped into the edgeless Corbino geometry, as shown in the schematic diagram in Fig.~\ref{fig1}a ~\cite{Nandi2012exciton,Li2019pairing,Zeng2019,Polshyn2018corbino}. In this geometry, the Landau level (LL) filling of each graphene layer, $\nu_1$ and $\nu_2$, can be independently controlled by adjusting the voltage bias on the top and bottom gate electrodes.
Electrical contacts, labeled 1 through 4, are connected to the inner and outer edges of both graphene layers. Fig.~\ref{fig1}b shows signatures of an exciton condensate at a total filling of $-1$, which corresponds to the filling fraction of hole-doped Landau levels ~\cite{Kellogg2004,Tutuc2004counterflow,Li2017superfluid,Liu2017superfluid,Liu2022crossover,Zeng2023solid,Nandi2012exciton,Eisenstein2014}. 
In the parallel flow (PF) geometry (top panel in Fig.~\ref{fig1}b), the excitonic state (highlighted by the blue-shaded stripe) behaves as an insulator. In contrast, this state exhibits non-zero bulk conductance in the counterflow geometry (middle panel). 

The most defining transport response is observed in the drag geometry (bottom panel), where the graphene layers are divided into two separate electrical circuits. In the active circuit, a drive current $I_{\text{drive}}$ is sent through the upper layer, causing charges of a specific sign to move in a given direction. Due to exciton pairing, carriers in the bottom layer are dragged in the same direction as those in the upper layer, generating a drag current $I_{\text{drag}}$ in the passive circuit. Since excitons consist of oppositely charged constituents, $I_{\text{drive}}$ and $I_{\text{drag}}$ flow in opposite directions but share the same amplitude. As a result, the exciton condensate, described by the (111) wavefunction ~\cite{Halperin1983}, acts like a perfect transformer, converting $100\%$ of the current in the drive layer into the drag circuit.  
As the d.c. voltage bias and temperature increase, the drag ratio deviates from unity, with the the drive current $I_{\text{drive}}$ becoming larger than the drag current $I_{\text{drag}}$ (see Fig.~\ref{fig1}c). This behavior indicates the existence of a critical threshold of d.c. voltage bias and temperature, beyond which the generation of unpaired charge carriers lead to the breakdown of the perfect drag condition.

We then employ the counterflow drag measurement to explore the low-temperature phase space of a quantum Hall graphene bilayer, defined by total filling $\nu_{\text{total}}=\nu_1+\nu_2$ and layer imbalance $\Delta \nu =\nu_1-\nu_2$. Fig.~\ref{fig1}d plots conductance measured from the PF geometry, $G_{\text{PF}}$, which probes the charge gaps associated with FQHE states. 
In Fig.~\ref{fig1}d, the white solid line outlines the region of the phase space where charge carriers in both graphene layers occupy the lowest LL, in the range of $-1 < \nu_1 <0 $ and $-1 < \nu_2 <0 $. For simplicity, we label the LL filling in this regime using the minority carrier filling $\tilde{\nu}_1=1+\nu_1$ and $\tilde{\nu}_2=1+\nu_2$. According to this convention, $\tilde{\nu}_{\text{total}} = \tilde{\nu}_1 + \tilde{\nu}_2$ and $\Delta \nu = \tilde\nu_1-\tilde\nu_2$. At $\tilde{\nu}_{\text{total}}=1$, the exciton condensate is shown as a prominent insulator with vanishing $G_{\text{PF}}$, which is depicted as dark blue in the chosen color scale.  Away from the total filling of one, insulating features form a well-defined two-dimensional (2D) pattern, pointing to a sequence of FQHE states with two-component correlation ~\cite{Li2019pairing}.

According to the composite fermion (CF) model, the two-component nature of the Coulomb interaction in a quantum Hall bilayer is modeled by forming CFs with intra- and interlayer magnetic flux attachments ~\cite{Jain.03,Jain2015CF,Eisenstein1990FQHE,Li2019pairing}. Notably, fractional excitons are directly linked to composite fermions constructions with interlayer flux attachment.  For simplicity, we denote a CF construction with $a$ intralayer and $b$ interlayer flux attachments as $^a_b$CF.
Effective fillings of CFs in each layer is defined as,
\begin{equation}\label{Eq:CF}
    {}^a_b\nu_1^{\ast} = \frac{\nu_1}{1-a\nu_1-b\nu_2}, \ \ {}^a_b\nu_2^{\ast} = \frac{\nu_2}{1-a\nu_2-b\nu_1}.
\end{equation}
Here, ${}^a_b\nu_i^{\ast}$ denotes effective filling $\nu^{\ast}$ of ${}^a_b$CFs on layer $i$. According to the CF model, a FQHE with a robust charge gap arises when ${}^a_b\nu^{\ast}_1$ and ${}^a_b\nu^{\ast}_2$ both take integer values ~\cite{Li2019pairing,Liu2019interlayer}. When ${}^a_b\nu^{\ast}_1$ is held at a constant integer value, varying ${}^a_b\nu^{\ast}_2$ stabilizes a sequence of FQHE state belonging to the $^a_b$CF construction. As described by Eq. (\ref{Eq:CF}), this sequence traces clear trajectories in the 2D phase space. For instance, the expected trajectories of \CFn, \CFi\, and \CFii\ states are shown as solid lines in Fig.~\ref{fig1}e-g and dashed lines in Fig.~\ref{fig1}d. These trajectories serves as a road map, outlining a systematic approach to understanding the connection between CFs and exciton pairing in quantum Hall bilayers (see Fig.~\ref{CFmap})  ~\cite{Li2019pairing,Scarola2001,Jain.03}. 

In this work, we utilize the counterflow drag measurement to identify two classes of fractional excitons, which are directly associated with the \CFi\ and \CFii\ constructions. The \CFn\ construction, on the other hand, lacks interlayer flux attachments, resulting in a different type of interlayer correlation for the corresponding FQHE state. This is beyond the scope of the current study and will not be explored here.

\begin{figure*}
\includegraphics[width=1\linewidth]{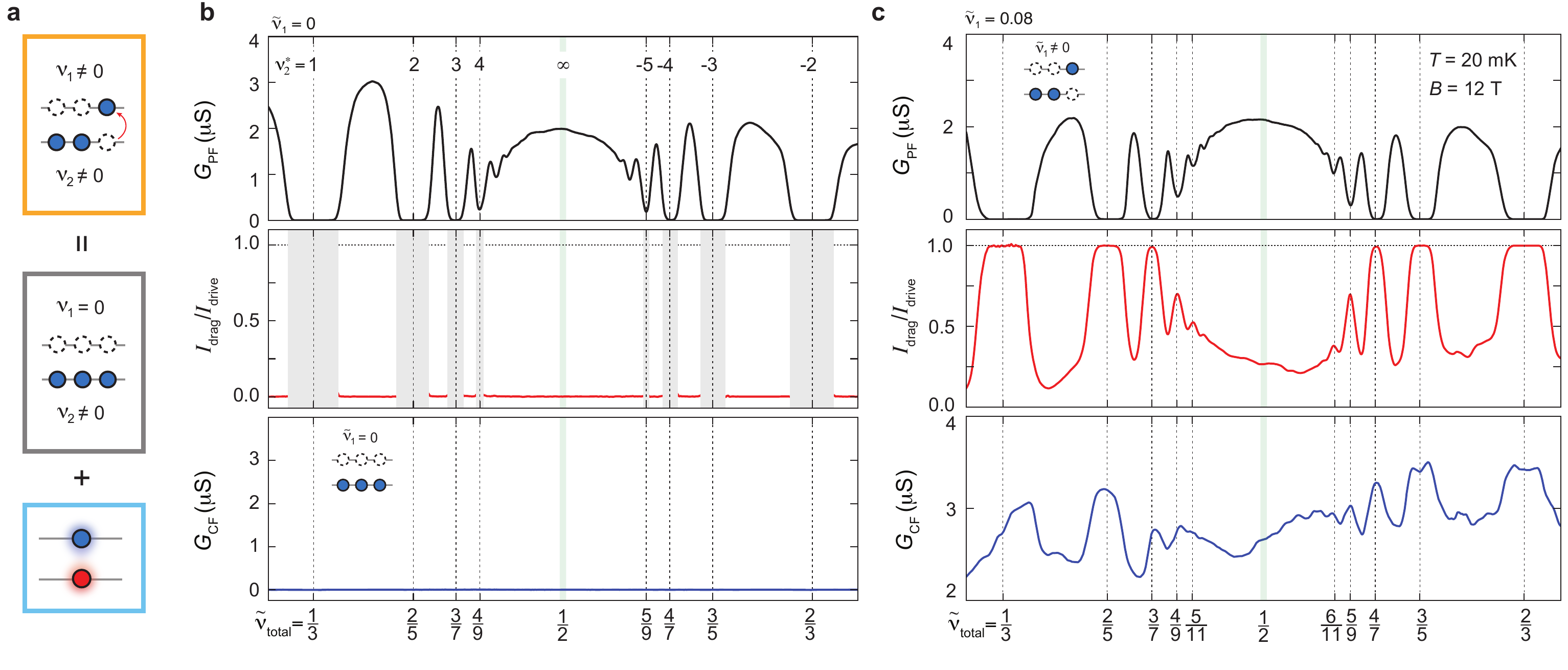}
\caption{\label{fig2}{\bf{A new sequence of FQHE defined by interlayer excitons.}} (a) Schematic diagram illustrating the introduction of interlayer excitons from a single-layer FQHE state by varying layer imbalance \dnu. (b-c) Parallel flow conductance $G_{\text{PF}}$ (top panel), drag ratio (middle panel), and counterflow conductance $G_{\text{CF}}$ (bottom panel) measured in (b) the single-layer regime at $\tilde{\nu}_1=0$ and (c) the two-component regime at $\tilde{\nu}_1=0.08$. In the middle panel of (b), drag ratio is ill defined at $^2_0\nu_2^{\ast}=N \in \mathbb{Z}$ as $I_{\text{drive}}$ and $I_{\text{drag}}$ are both zero. These locations are masked by gray-shaded stripes. 
}
\end{figure*}

\begin{figure*}
\includegraphics[width=0.99\linewidth]{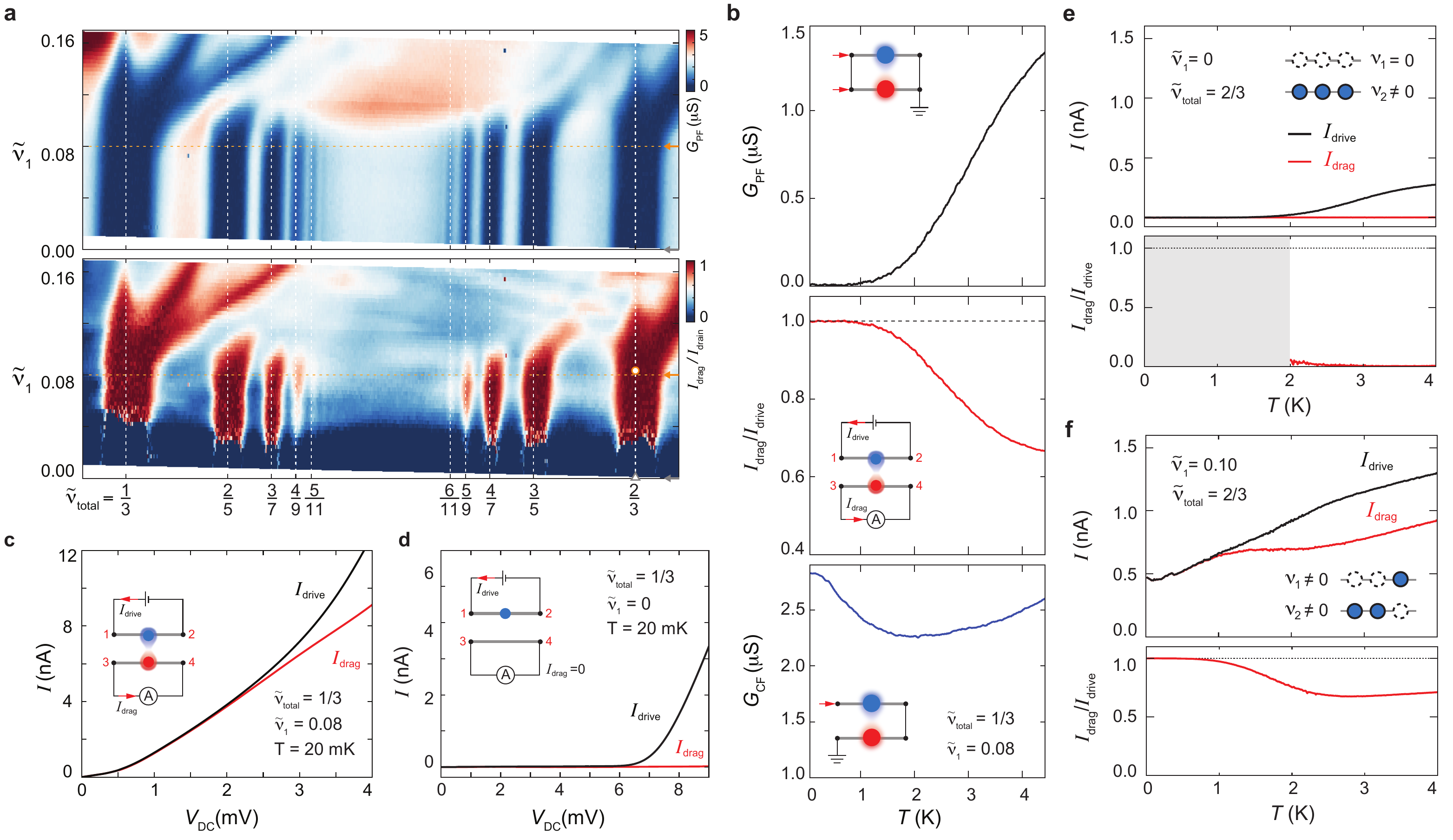}
\caption{\label{fig3}{\bf{Constructing a sequence of FQHE with \CFii\ excitons.}} (a) Parallel flow conductance $G_{\text{PF}}$ (top panel) and drag ratio (bottom panel) as a function of $\tilde{\nu}_{\text{total}}$ and $\tilde{\nu}_1$. Horizontal line traces marked by gray and orange arrows correspond to the measurements in Fig.~\ref{fig2}b and c.  (b) $G_{\text{PF}}$ (top panel), drag ratio $I_{\text{drag}}/I_{\text{drive}}$ (middle panel), and $G_{\text{CF}}$ (bottom panel) as a function of $T$ measured at $\tilde{\nu}_{\text{total}} = 1/3$ and $\tilde\nu_1 = 0.08$.   (c-d) Drive current $I_{\text{drive}}$ and drag current $I_{\text{drag}}$ as a function of d.c. voltage bias, measured for (c) the excitonic state at $\tilde\nu_1 = 0.08$ and (d) in the single-layer regime at $\tilde\nu_1 = 0$.  (e-f) Drive current $I_{\text{drive}}$  and dragcurrent $I_{\text{drag}}$ as a function of temperature, measured with the drag geometry at $\tilde{\nu}_{\text{total}} = 2/3$ and (e) in the single-layer regime at $\tilde{\nu}_1 = 0$ and (f) in the 2-component regime at $\tilde\nu_1 = 0.10$ (corresponding to open gray triangle and open orange circle in (d), respectively).
}
\end{figure*}

\begin{figure*}
\includegraphics[width=1\linewidth]{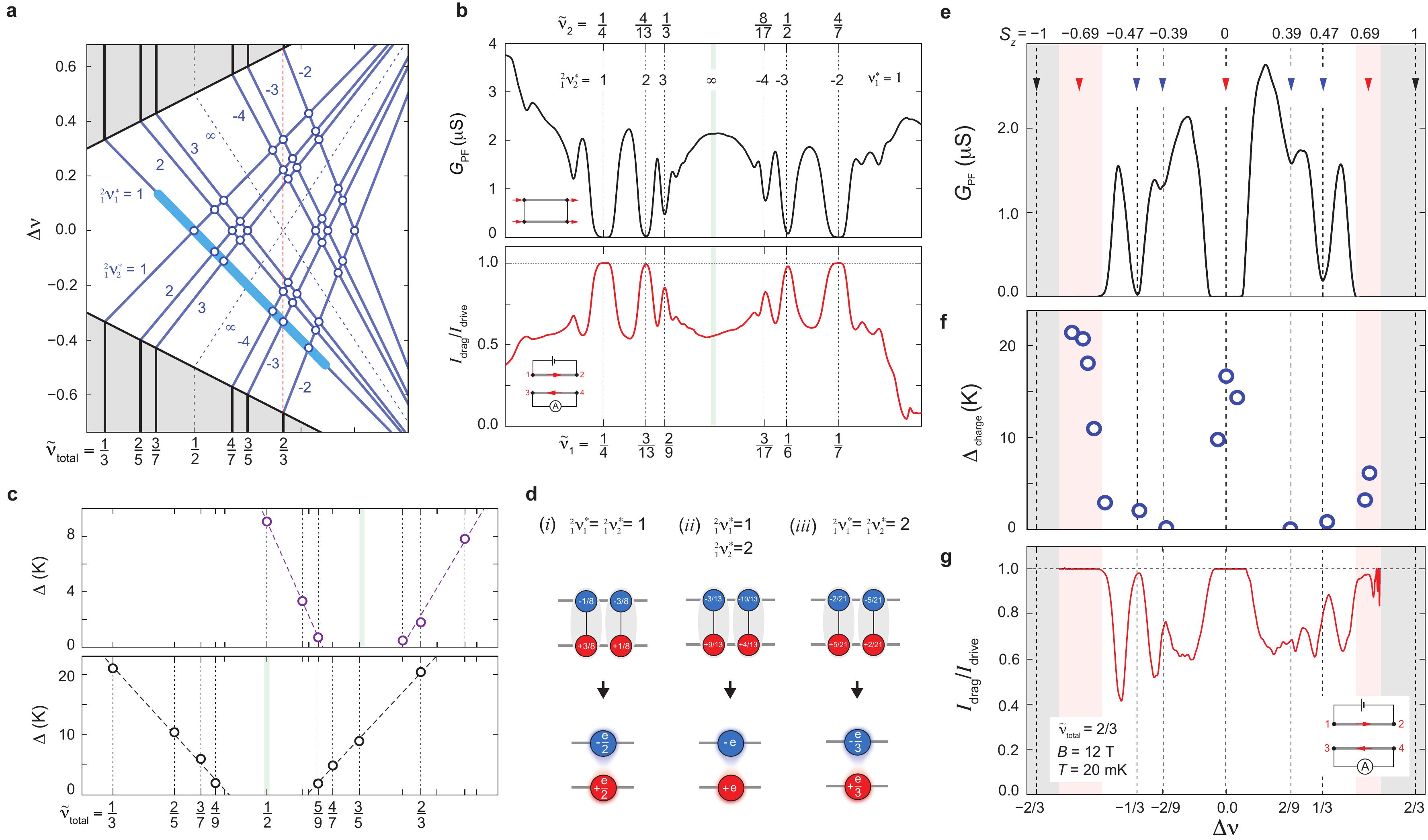}
\caption{\label{fig4}{\bf{Fractional excitons from the \CFi\ sequence.} } 
(a) Schematic diagram of \CFi\ states across a portion of $\tilde{\nu}_{\text{total}}$-\dnu\ map. Blue solid lines denote integer values of $^2_1\nu_{i}^{\ast}$. Vertical red dashed line marks the expected trajectory of the \CFii\ exciton order at $\tilde{\nu}_{\text{total}} = 2/3$.  (b) $G_{\text{PF}}$ (top), and drag ratio (bottom) measured along the light blue solid line in panel (a). (c) Energy gap extracted from the activation behavior in $G_{\text{PF}}$ for the \CFi sequence (top) and \CFn\ (bottom) sequence of FQHE states. (d) Schematic diagram shows excitonic pairing between quasiparticle and quasi-hole excitations for different \CFi states. (e) $G_{\text{PF}}$, (f) charge gap $\Delta$, and (g) drag ratio $I_{\text{drag}}/I_{\text{drive}}$ as a function of \dnu\ measured at $\tilde{\nu}_{\text{total}}= 2/3$. Vertical arrows mark $9$ different ground states tunable with layer pseudo-spin polarization $S_z$. Grey and red shaded background mark \dnu\ ranges that correspond to the single-layer regime and the \CFii\ exciton states.}
\end{figure*}

The first class of fractional exciton emerges with a sequence of two-component FQHE states, which is stabilized by adding excitonic pairing to the Jain-sequence of FQHE.  As shown in Fig.~\ref{fig2}a, varying \dnu\ in a single-layer FQHE effectively generates interlayer excitons. 
Fig.~\ref{fig2}b displays the transport signatures of the Jain sequence of FQHE states. As a function of varying $\tilde{\nu}_{\text{total}}$ at $\tilde{\nu}_1 = 0$, a sequence of insulators is revealed by vanishing $G_{\text{PF}}$, which appears at ${}^2_0\nu_2^{\ast}=N\in \mathbb{Z}$  (vertical dashed lines in Fig.~\ref{fig2}b). These FQHE states exhibit vanishing drag response (middle panel of Fig.~\ref{fig2}b) and zero counterflow conductance $G_{\text{CF}} = 0$ (bottom panel of Fig.~\ref{fig2}b). This confirms the lack of interlayer correlation in the single-layer regime, where charge carriers are confined to just one layer.

From this Jain-sequence of FQHE, exciton pairing is induced by adjusting $\tilde{\nu}_1$ to a non-zero value. 
Fig.~\ref{fig2}c plots the transport response as a function of varying $\tilde{\nu}_{\text{total}}$ at $\tilde{\nu}_1 = 0.08$. Along this line, a series of FQHE states appears at $\tilde\nu_{\text{total}} = N/(1+2N)$, with $N\in \mathbb{Z}$, which is revealed by vanishing bulk conductance in the PF geometry, $G_{\text{PF}}=0$. Two key observations provide unambiguous evidence of excitonic pairing. First, each insulator with vanishing $G_{\text{PF}}$ is accompanied by the perfect drag ratio $I_{\text{drag}}/I_{\text{drive}} = 1$ in the drag geometry (middle panel Fig.~\ref{fig2}c). Secondly, FQHE states are highly conductive in the counterflow geometry, where local maxima in $G_{\text{CF}}$ coincide with vanishing $G_{\text{PF}}$ (bottom panel Fig.~\ref{fig2}c). Together, these transport responses indicate the coexistence of a robust FQHE energy gap and the charge-neutral mode of interlayer exciton. While the location of these FQHE states in $\tilde\nu_{\text{total}}$ aligns with the conventional Jain sequence, the presence of exciton pairing incorporates interlayer correlation into the wavefunction. This distinguishes them from the Jain sequence and underscores the unique interplay between excitonic pairing and the FQHE, resulting in a distinct two-component state.

To further explore the connection between the Jain sequence and the fractional exciton states, Fig.~\ref{fig3}a plots the evolution of transport responses with varying $\tilde{\nu}_1$. Fig.~\ref{fig2}b and c correspond to line traces of the maps in Fig.~\ref{fig3}a, taken along $\tilde{\nu}_1 = 0$ and $\tilde{\nu}_1 = 0.08$, respectively. The evolution with varying $\tilde{\nu}_1$ reveals two key aspects regarding the construction of the fractional excitons. First, the charge gap of each Jain sequence state remains robust despite the introduction of excitons, as indicated by the vanishing $G_{\text{PF}}$ with varying \dnu\ (the top panel of Fig.~\ref{fig3}a). This observation is consistent with the charge neutral nature of interlayer excitons, which do not affect the overall charge gap of the FQHE state. Second, this class of fractional excitons extends along lines of constant $\tilde{\nu}_{\text{total}}$ over a range of layer imbalance \dnu.  As shown in the bottom panel of Fig.~\ref{fig3}a, the perfect drag response, shown as red in the chosen color scale, extends along a vertical trajectory. Such robustness against \dnu\ resembles the behavior observed in the $(111)$ state ~\cite{Li2017superfluid,Liu2017superfluid,Zeng2023solid}, offering a crucial clue to the origin of fractional excitons.

Similar to the $(111)$ state, which occurs at an integer total Landau level filling fraction across two graphene layers, the first class of fractional excitons emerges when the total effective filling of \CFii\ reaches integer values,  ${}^2_2\nu_{\text{total}}^{\ast} = {}^2_2\nu_1^{\ast} + {}^2_2\nu_2^{\ast} =N$, with $N\in \mathbb{Z}$. This condition, represented by the vertical red trajectories shown in Fig.~\ref{fig1}g, suggests that the fractional excitons emerge from exciton pairing between partially filled $\Lambda$-levels of \CFii.  The direct analogy with the $(111)$ state implies that the FQHE in this regime exhibits a similar form of interlayer correlation. Consequently, the resulting excitons are expected to obey bosonic statistics, with the low-temperature ground state described by a Bose-Einstein condensate.

A neutral superfluid phase has been predicted to appear at total filling $1/3$ in a quantum Hall bilayer, described by a $(333)$ wavefunction. This corresponds to the fractional exciton state at $\tilde{\nu}_{\text{total}} = 1/3$ in Fig.~\ref{fig2}c and Fig.~\ref{fig3}a. Although Corbino measurements are inherently two-terminal and cannot directly identify dissipationless exciton flow, we can still gain insight by analyzing the temperature dependence and I-V characteristics of the (333) state in comparison with those observed in the (111) state. 
Fig.~\ref{fig3}b plots the temperature dependence of transport signatures at $\tilde{\nu}_{\text{total}}= 1/3$. Above a critical temperature around $T = 1$ K, hallmark signatures of excitonic transport begin to diminish: $G_{\text{PF}}$ increases from zero and the drag ratio deviates from one as $T$ rises. This $T$-dependence suggests that free charges are generated at $T > 1$ K, resulting in the formation of an additional parallel transport channel that operates independently of exciton flow. Furthermore, the reduction in counterflow conductance $G_{\text{CF}}$ indicates that exciton flow is increasingly suppressed with rising $T$.

Theoretical work has recognized that an excitonic state described by the $(nnn)$ wavefunction exhibits collective excitations in the form of unpaired vortexes, known as merons and anti-merons  ~\cite{Wen1992,Moon1995,Huse2005,Kellogg2005,Eisenstein2014,Liu2022crossover}. Since these unpaired vortexes carry a non-zero electric charge, the observed vanishing $G_{\text{PF}}$ and perfect drag response suggest that all merons and anti-merons form charge-netural form pairs at low temperature. According to the Berezinskii-Kosterlitz-Thouless (BKT) model ~\cite{KT1973,Berezinskii1972BKT}, this vortex pairing process defines the transition into a low-temperature condensate phase at $\tilde{\nu}_{\text{total}}= 1/3$.

The potential condensate is further supported by the I-V characteristics measured from the drag geometry at $\tilde{\nu}_{\text{total}}= 1/3$ and $\tilde{\nu}_1 = 0.08$ (Fig.~\ref{fig3}c). Below a critical d.c. voltage bias of $2$ mV, the FQHE state acts like a perfect transformer, converting $100\%$ of the current in the drive layer into the drag circuit.  This is in excellent agreement with the observed behavior of the $(111)$ state in Fig.~\ref{fig1}c. 
In contrast, the FQHE at $\tilde{\nu}_{\text{total}}= 1/3$ and $\tilde{\nu}_1 = 0$, which is described by the single-layer Laughlin wavefunction, acts like a decoupler. Above a threshold in the d.c. voltage bias, the onset in $I_{\text{drive}}$ is accompanied by a vanishingly small drag current. This stark difference between the two FQHE states highlights the unique nature of the fractional exciton state and its capacity to facilitate a perfect drag response. This observation supports the hypothesis of a low-temperature excitonic condensate phase at $\tilde{\nu}_{\text{total}} = 1/3$.

Along the same veins, Fig.~\ref{fig3}e-f examines the fractional exciton order at $\tilde{\nu}_{\text{total}}= 2/3$. 
At $\tilde{\nu}_1=0$, the drive and drag circuits are decoupled by the lack of exciton flow. In this regime, the FQHE in layer 2 arises from the intralayer Coulomb correlation, with thermal excitations also confined to a single layer. However, tuning layer 1 to $\tilde{\nu}_1 = 0.10$ induces exciton pairing, which stabilizes the perfect drag response at $T < 1$ K. Furthermore, the presence of the fractional exciton maintains a nonzero counterflow drag response down to the base temperature of the dilution fridge, suggesting that the excitons enable a gapless charge-neutral mode.

Next, we turn our attention to the \CFi\ sequence of FQHE, which hosts a different type of excitons with non-bosonic properties.
According to Eq.~1, the \CFi\ sequence of incompressible states are expected at points where both ${}^2_1\nu_{1}^{\ast}$ and ${}^2_1\nu_{2}^{\ast}$ are integer values, indicated by blue open circles in Fig.~\ref{fig4}a. 
Transport measurements along the light blue solid line in Fig.~\ref{fig4}a, which is defined by varying ${}^2_1\nu_2^{\ast}$ while holding ${}^2_1\nu_1^{\ast}$ constant, offers unambiguous support for the \CFi\ construction. As shown in the top panel of Fig.~\ref{fig4}b, a sequence of FQHE states with vanishing $G_{\text{PF}}$ appears at integer values of ${}^2_1\nu_{2}^{\ast}$ (indicated by vertical dashed lines). By examining the thermally activated behavior of $G_{\text{PF}}$, we extract the value of the charge gap $\Delta$, which demonstrates a characteristic hierarchical behavior (Fig.~\ref{fig4}c): the charge gap becomes more robust with decreasing effective filling and less robust with increasing effective filling. This hierarchical behavior, combined with the precise locations of the FQHE sequence, provides strong evidence that the \CFi\ construction is responsible for the formation of the charge gap. 

Remarkably, the \CFi\ sequence of FQHE states are accompanied by a perfect drag response, as shown in the bottom panel of Fig.~\ref{fig4}b. This provides a clear indication of the presence of exciton pairing. However, it is important to note that the CF model does not inherently provide a mechanism for exciton pairing in the ground state of the \CFi\ construction. Nevertheless, the FQHE wavefunctions at these states, located at integer values of ${}^2_1\nu^*_1$ and ${}^2_1\nu^*_2$, can be analyzed using the ${\bf K}$-matrices formalism \cite{Wen2004quantum}.
By extracting the ${\bf K}$-matrices, we can gain further insight into the likely composition of the interlayer excitons  (see Section III in Method).
In the presence of a FQHE charge gap, exciton formation occurs naturally by pairing quasiparticle and quasihole excitations. At ${}^2_1\nu^*_1={}^2_1\nu^*_2=1$, the FQHE state is described by the $(331)$ wavefunction, which enables a quasiparticle with charge $-e/8$ on one layer and $+3e/8$ on the other, and a quasihole with charge $-3e/8$ and $+e/8$ across the two layers, as illustrated in panel (i) of Fig.~\ref{fig4}d.  
A bound state of such a quasiparticle and quasihole leads to the formation of a fractional exciton with particle and hole charge of $\pm e/2$. Remarkably, this exciton is not a composite boson but instead obeys fermionic statistics (see Section III in Method). 
The emergence of a fermionic exciton at half filling of a two-component system has been previously proposed in theoretical discussions ~\cite{Barkeshi2018}.

In a similar fashion, the exciton composition at ${}^2_1\nu^*_1=1$ and ${}^2_1\nu^*_2=2$ involves a quasiparticle with charge $+9e/13$ in one layer and $-3e/13$ in the other, and a quasihole with charge $+4e/13$ and $-10e/13$ across the two layers. This pairing results in an exciton with a net charge of $\pm e$. The two types of quasiparticles that form an exciton have different charges due to the presence of layer asymmetry.
While the resulting exciton has $\pm e$ charge across two layers, its depairing leads to fractionally charged quasiparticle and quasihole, as shown in panel (ii) of Fig.~\ref{fig4}d. Analysis shows that excitons with this composition obey bosonic statistics. 

Following the same ${\bf K}$-matrix analysis, further increasing the effective filling of \CFi\ reveals fractional excitons that exhibit anyonic behavior. 
At ${}^2_1\nu_1^{\ast} = {}^2_1\nu_2^{\ast} = 2$, exciton pairing gives rise to a bound state with $\pm e/3$ charge across two layers, as shown in panel (iii) of Fig.~\ref{fig4}d. This exciton acquires a statistical phase of $4\pi/3$ upon exchange.

The possibility of fractional excitons with non-bosonic statistics raises intriguing questions regarding the nature of the low-temperature phase. This is further complicated by the construction of pairing between quasiparticles and quasiholes. Since quasiparticles carry a non-zero energy cost, their population is expected to vanish as temperature approaches zero. However, our observation of a perfect drag ratio at $T = 20$ mK, alongside a charge gap of $\Delta \sim 8$ K, suggests that the formation of charge-neutral exciton substantially reduces the energy cost of quasiparticle excitations. This points to non-bosonic excitons as a low-energy neutral mode within the two-component FQHE. It is also worth considering that a few fractional excitons could potentially combine to form a larger, bosonic composite particle. Although the intricate composition of such a construction makes this scenario unlikely, it cannot be definitively ruled out based solely on transport measurements. Therefore, our findings raise an important question for future research efforts. 

Unlike \CFii\ excitons that extends along lines of constant $\tilde{\nu}_{\text{total}}$, \CFi\ excitons are only stable around specific points in the phase space, defined by integer values of ${}^2_1\nu^*_1$ and ${}^2_1\nu^*_2$.
This distinction is demonstrated by a series of transitions at $\tilde{\nu}_{\text{total}}= 2/3$ as a function of varying \dnu. At this filling, changing \dnu\ stabilizes a cascade of FQHE states with robust energy gaps (indicated by arrows and vertical dashed lines in Fig.~\ref{fig4}e-g). The construction of each FQHE is highlighted by the schematic map in Fig.~\ref{fig4}a, where $\tilde{\nu}_{\text{total}}= 2/3$ is marked by the vertical red dashed line. With full layer polarization at \dnu $ = \pm 2/3$, the single-layer regime is highlighted by a grey-shaded background, where a lack of interlayer correlation gives rise to a FQHE state with the \CFn\ construction. As \dnu\ deviates from full polarization, \CFii\ excitons occupy a range of \dnu\ marked with a red-shaded background. Further decreasing \dnu\ stabilizes \CFi\ exciton states, specifically at (${}^2_1\nu_1^{\ast}=1$, ${}^2_1\nu_2^{\ast}=-3$), and (${}^2_1\nu_1^{\ast}=2$, ${}^2_1\nu_2^{\ast}=-4$). These points are uniquely connected to the \CFi\ construction, setting them apart from the more continuous presence of \CFii\ excitons.

In quantum Hall bilayers, adjusting layer imbalance \dnu\ allows access to the full range of pseudo-spin polarization $S_z$, which is defined as,
\begin{equation}
    S_z = \frac{\tilde{\nu}_1-\tilde{\nu}_2}{\tilde{\nu}_1+\tilde{\nu}_2} = \frac{\Delta \nu}{\tilde{\nu}_{\text{total}}}. 
\end{equation}   
Given the experimental challenges to control and probe electronic quantum numbers,  such as spin, valley isospin, and sublattice, layer pseudo-spin in quantum Hall bilayers offers an exceptional simulator for these quantum numbers. Although the detection of neutral modes presents an outstanding challenge for most two-component electron systems ~\cite{review-edge,Heiblum.10}, quantum Hall bilayer are an exception, as excitonic neutral modes can be easily detected and examined with counterflow measurements.

Using the language of layer pseudo-spin, the presence of a robust charge gap at $S_z=0$ suggests a pseudo-spin antiferromagnetic insulator (see section IV in Methods). This insulator is accompanied by a gapless neutral mode, as evidenced by the observation of the perfect drag ratio (see Fig.~\ref{fig4}g). This phenomenon is directly comparable to the spin superfluid in a single 2D layer, which has been explored in association with a spin antiferromagnetic insulator ~\cite{Takei2014Spinsuperfluid,Takei2016Spinsuperfluid,Wei2018spinwave,Zhou2022spinwave}. This analogy underscores the unique capability of quantum Hall bilayers to simulate and explore complex electronic behaviors in the FQHE landscape, offering unprecedented insights into phenomena such as antiferromagnetic and canted-antiferromagnetic orders, as well as neutral mode dynamics, which are otherwise challenging to detect and study in conventional 2D electron systems.

In summary, our observations underscores the pivotal role of charge neutral modes, including but not limited to interlayer excitons, in defining the electronic orders across the FQHE landscape. Given this critical role of excitons, our findings are poised to have a far-reaching impact beyond the conventional quantum Hall effect regime. For example, the involvement of multiple electron species and neutral modes may be key to understanding the anomalous version of FQHE recently discovered in various van der Waals structures under zero magnetic field  ~\cite{Cai2023,Park2023,Xu2023,Zeng2023,Lu2024}. 
Overall, the studies of excitons in the FQHE unlock a vast array of research opportunities, promising to attract widespread interest and stimulate further exploration into the intricate behavior of multi-component electronic systems.

\section*{acknowledgments}
J.I.A.L. wishes to acknowledge helpful discussion with Jainendra Jain. This material is based on the work supported by the Air Force Office of Scientific Research under award no. FA9550-23-1-0482. N.J.Z., R.Q.N., and J.I.A.L. acknowledge support from the Air Force Office of Scientific Research. N.J.Z. acknowledges partial support from the Jun-Qi fellowship. R.Q.N. and J.I.A.L. acknowledges partial support from the National Science Foundation EPSCoR Program under NSF Award OIA-2327206. Any opinions, findings and conclusions or recommendations expressed in this material are those of the author(s) and do not necessarily reflect those of the National Science Foundation. This work was performed in part at Aspen Center for Physics, which is supported by National Science Foundation grant PHY-2210452. A portion of this work was performed at the National High Magnetic Field Laboratory, which is supported by National Science Foundation Cooperative Agreement No. DMR-1157490 and the State of Florida. N.B. and D.E.F. were supported in part by the National Science Foundation under Grant No. DMR-2204635. K.W. and T.T. acknowledge support from the JSPS KAKENHI (Grant Numbers 21H05233 and 23H02052) and World Premier International Research Center Initiative (WPI), MEXT, Japan. 

\section*{Data availability}
The data that support the plots within this paper and other findings of this study are available from the corresponding author upon reasonable request.

\section*{Author contribution}
N.J.Z. and J.I.A.L. conceived the project. N.J.Z., R.Q.N., and
X.L. fabricated the device. N.J.Z. and R.Q.N. performed the measurement. N.B. and D.E.F. provided theoretical inputs. K.W. and T.T. provided the material. N.J.Z., R.Q.N., N.B., D.E.F., and J.I.A.L wrote the manuscript together.

\section*{Competing financial interests}
The authors declare no competing financial interests.

\bibliography{Li_ref_ExcitonManuscript}

\begin{thebibliography}{10}
\expandafter\ifx\csname url\endcsname\relax
  \def\url#1{\texttt{#1}}\fi
\expandafter\ifx\csname urlprefix\endcsname\relax\def\urlprefix{URL }\fi
\providecommand{\bibinfo}[2]{#2}
\providecommand{\eprint}[2][]{\url{#2}}

\bibitem{review-FH}
\bibinfo{author}{Feldman, D.~E.} \& \bibinfo{author}{Halperin, B.~I.}
\newblock \bibinfo{title}{Fractional charge and fractional statistics in the quantum {H}all effects}.
\newblock \emph{\bibinfo{journal}{Rep. Prog. Phys.}} \textbf{\bibinfo{volume}{84}}, \bibinfo{pages}{076501} (\bibinfo{year}{2021}).

\bibitem{Halperin1983}
\bibinfo{author}{Halperin, B.~I.}
\newblock \bibinfo{title}{Theory of the quantized {H}all conductance}.
\newblock \emph{\bibinfo{journal}{Helv. Phys. Acta}} \textbf{\bibinfo{volume}{56}}, \bibinfo{pages}{75--102} (\bibinfo{year}{1983}).

\bibitem{Yang.94}
\bibinfo{author}{Yang, K.} \emph{et~al.}
\newblock \bibinfo{title}{Quantum ferromagnetism and phase transitions in double-layer quantum {H}all systems}.
\newblock \emph{\bibinfo{journal}{Phys. Rev. Lett.}} \textbf{\bibinfo{volume}{72}}, \bibinfo{pages}{732--735} (\bibinfo{year}{1994}).

\bibitem{Moon1995}
\bibinfo{author}{Moon, K.} \emph{et~al.}
\newblock \bibinfo{title}{Spontaneous interlayer coherence in double-layer quantum {H}all systems: Charged vortices and {K}osterlitz-{T}houless phase transitions}.
\newblock \emph{\bibinfo{journal}{Phys. Rev. B}} \textbf{\bibinfo{volume}{51}}, \bibinfo{pages}{5138} (\bibinfo{year}{1995}).

\bibitem{Girvin1996multi}
\bibinfo{author}{Girvin, S.} \& \bibinfo{author}{MacDonald, A.}
\newblock \bibinfo{title}{Multicomponent quantum {H}all systems: The sum of their parts and more}.
\newblock \emph{\bibinfo{journal}{Perspectives in Quantum Hall Effects: Novel Quantum Liquids in Low-Dimensional Semiconductor Structures}} \bibinfo{pages}{161--224} (\bibinfo{year}{1996}).

\bibitem{Laughlin1984exciton}
\bibinfo{author}{Laughlin, R.}
\newblock \bibinfo{title}{Excitons in the fractional quantum {H}all effect}.
\newblock \emph{\bibinfo{journal}{Physica B+C}} \textbf{\bibinfo{volume}{126}}, \bibinfo{pages}{254--259} (\bibinfo{year}{1984}).

\bibitem{Wen1992}
\bibinfo{author}{Wen, X.-G.} \& \bibinfo{author}{Zee, A.}
\newblock \bibinfo{title}{Neutral superfluid modes and ``magnetic'' monopoles in multilayered quantum {H}all systems}.
\newblock \emph{\bibinfo{journal}{Phys. Rev. Lett.}} \textbf{\bibinfo{volume}{69}}, \bibinfo{pages}{1811--1814} (\bibinfo{year}{1992}).

\bibitem{Kamilla1997exciton}
\bibinfo{author}{Kamilla, R.~K.} \& \bibinfo{author}{Jain, J.}
\newblock \bibinfo{title}{Excitonic instability and termination of fractional quantum {H}all effect}.
\newblock \emph{\bibinfo{journal}{Phys. Rev. B}} \textbf{\bibinfo{volume}{55}}, \bibinfo{pages}{R13417} (\bibinfo{year}{1997}).

\bibitem{Yang1994quasiexciton}
\bibinfo{author}{Yang, J.}
\newblock \bibinfo{title}{Quasiexcitons in the fractional quantum {H}all effect}.
\newblock \emph{\bibinfo{journal}{Phys. Rev. B}} \textbf{\bibinfo{volume}{49}}, \bibinfo{pages}{5443--5447} (\bibinfo{year}{1994}).

\bibitem{Park2000fractionalexciton}
\bibinfo{author}{Park, K.} \& \bibinfo{author}{Jain, J.~K.}
\newblock \bibinfo{title}{Two-roton bound state in the fractional quantum {H}all effect}.
\newblock \emph{\bibinfo{journal}{Phys. Rev. Lett.}} \textbf{\bibinfo{volume}{84}}, \bibinfo{pages}{5576--5579} (\bibinfo{year}{2000}).

\bibitem{Park2001fractionalexciton}
\bibinfo{author}{Park, K.}
\newblock \bibinfo{title}{Charged excitons of composite fermions in the fractional quantum {H}all effect}.
\newblock \emph{\bibinfo{journal}{Solid State Commun.}} \textbf{\bibinfo{volume}{121}}, \bibinfo{pages}{19--23} (\bibinfo{year}{2001}).

\bibitem{Jain2005fractionalexciton}
\bibinfo{author}{Jain, J.~K.}, \bibinfo{author}{Park, K.}, \bibinfo{author}{Peterson, M.~R.} \& \bibinfo{author}{Scarola, V.~W.}
\newblock \bibinfo{title}{Composite fermion theory of excitations in the fractional quantum {H}all effect}.
\newblock \emph{\bibinfo{journal}{Solid State Commun.}} \textbf{\bibinfo{volume}{135}}, \bibinfo{pages}{602--609} (\bibinfo{year}{2005}).

\bibitem{Quinn2006quasiexciton}
\bibinfo{author}{Quinn, J.~J.}, \bibinfo{author}{W{\'o}js, A.} \& \bibinfo{author}{G{\l}adysiewicz, A.}
\newblock \bibinfo{title}{Fractional quasiexcitons in incompressible electron liquids}.
\newblock \emph{\bibinfo{journal}{Physica E}} \textbf{\bibinfo{volume}{34}}, \bibinfo{pages}{280--283} (\bibinfo{year}{2006}).

\bibitem{Barkeshi2018}
\bibinfo{author}{Barkeshli, M.}, \bibinfo{author}{Nayak, C.}, \bibinfo{author}{Papi\ifmmode~\acute{c}\else \'{c}\fi{}, Z.}, \bibinfo{author}{Young, A.} \& \bibinfo{author}{Zaletel, M.}
\newblock \bibinfo{title}{Topological {E}xciton {F}ermi {S}urfaces in {T}wo-{C}omponent {F}ractional {Q}uantized {H}all {I}nsulators}.
\newblock \emph{\bibinfo{journal}{Phys. Rev. Lett.}} \textbf{\bibinfo{volume}{121}}, \bibinfo{pages}{026603} (\bibinfo{year}{2018}).

\bibitem{Zhang2023fractionalexciton}
\bibinfo{author}{Zhang, Y.-H.}, \bibinfo{author}{Zhu, Z.} \& \bibinfo{author}{Vishwanath, A.}
\newblock \bibinfo{title}{{XY*} {T}ransition and {E}xtraordinary {B}oundary {C}riticality from {F}ractional {E}xciton {C}ondensation in {Q}uantum {H}all {B}ilayer}.
\newblock \emph{\bibinfo{journal}{Phys. Rev. X}} \textbf{\bibinfo{volume}{13}}, \bibinfo{pages}{031023} (\bibinfo{year}{2023}).

\bibitem{Kwan2022quasiexciton}
\bibinfo{author}{Kwan, Y.~H.}, \bibinfo{author}{Hu, Y.}, \bibinfo{author}{Simon, S.~H.} \& \bibinfo{author}{Parameswaran, S.~A.}
\newblock \bibinfo{title}{Excitonic fractional quantum {H}all hierarchy in moir\'e heterostructures}.
\newblock \emph{\bibinfo{journal}{Phys. Rev. B}} \textbf{\bibinfo{volume}{105}}, \bibinfo{pages}{235121} (\bibinfo{year}{2022}).

\bibitem{faugno2020}
\bibinfo{author}{Faugno, W.~N.}, \bibinfo{author}{Balram, A.~C.}, \bibinfo{author}{W{\'o}js, A.} \& \bibinfo{author}{Jain, J.~K.}
\newblock \bibinfo{title}{Theoretical phase diagram of two-component composite fermions in double-layer graphene}.
\newblock \emph{\bibinfo{journal}{Phys. Rev. B}} \textbf{\bibinfo{volume}{101}}, \bibinfo{pages}{085412} (\bibinfo{year}{2020}).

\bibitem{Eisenstein2014}
\bibinfo{author}{Eisenstein, J.~P.}
\newblock \bibinfo{title}{Exciton {C}ondensation in {B}ilayer {Q}uantum {H}all {S}ystems}.
\newblock \emph{\bibinfo{journal}{Annu. Rev. of Condens. Matter Phys.}} \textbf{\bibinfo{volume}{5}}, \bibinfo{pages}{159--181} (\bibinfo{year}{2014}).

\bibitem{Li2017superfluid}
\bibinfo{author}{Li, J. I.~A.}, \bibinfo{author}{Taniguchi, T.}, \bibinfo{author}{Watanabe, K.}, \bibinfo{author}{Hone, J.} \& \bibinfo{author}{Dean, C.~R.}
\newblock \bibinfo{title}{Excitonic superfluid phase in double bilayer graphene}.
\newblock \emph{\bibinfo{journal}{Nat. Phys.}} \textbf{\bibinfo{volume}{13}}, \bibinfo{pages}{751--755} (\bibinfo{year}{2017}).

\bibitem{Liu2017superfluid}
\bibinfo{author}{Liu, X.}, \bibinfo{author}{Taniguchi, T.}, \bibinfo{author}{Watanabe, K.}, \bibinfo{author}{Halperin, B.} \& \bibinfo{author}{Kim, P.}
\newblock \bibinfo{title}{Quantum {H}all drag of exciton condensate in graphene}.
\newblock \emph{\bibinfo{journal}{Nat. Phys.}} \textbf{\bibinfo{volume}{13}}, \bibinfo{pages}{746--750} (\bibinfo{year}{2017}).

\bibitem{Liu2022crossover}
\bibinfo{author}{Liu, X.} \emph{et~al.}
\newblock \bibinfo{title}{Crossover between strongly coupled and weakly coupled exciton superfluids}.
\newblock \emph{\bibinfo{journal}{Science}} \textbf{\bibinfo{volume}{375}}, \bibinfo{pages}{205--209} (\bibinfo{year}{2022}).

\bibitem{SI}
\emph{\bibinfo{journal}{Please see the supplementary materials}} .

\bibitem{Nandi2012exciton}
\bibinfo{author}{Nandi, D.}, \bibinfo{author}{Finck, A. D.~K.}, \bibinfo{author}{Eisenstein, J.~P.}, \bibinfo{author}{Pfeiffer, L.~N.} \& \bibinfo{author}{West, K.~W.}
\newblock \bibinfo{title}{Exciton condensation and perfect coulomb drag}.
\newblock \emph{\bibinfo{journal}{Nature}} \textbf{\bibinfo{volume}{488}}, \bibinfo{pages}{481--484} (\bibinfo{year}{2012}).

\bibitem{Li2019pairing}
\bibinfo{author}{Li, J.} \emph{et~al.}
\newblock \bibinfo{title}{Pairing states of composite fermions in double-layer graphene}.
\newblock \emph{\bibinfo{journal}{Nat. Phys.}} \textbf{\bibinfo{volume}{15}}, \bibinfo{pages}{898--903} (\bibinfo{year}{2019}).

\bibitem{Zeng2019}
\bibinfo{author}{Zeng, Y.} \emph{et~al.}
\newblock \bibinfo{title}{High-{Q}uality {M}agnetotransport in {G}raphene {U}sing the {E}dge-{F}ree {C}orbino {G}eometry}.
\newblock \emph{\bibinfo{journal}{Phys. Rev. Lett.}} \textbf{\bibinfo{volume}{122}}, \bibinfo{pages}{137701} (\bibinfo{year}{2019}).

\bibitem{Polshyn2018corbino}
\bibinfo{author}{Polshyn, H.} \emph{et~al.}
\newblock \bibinfo{title}{Quantitative {T}ransport {M}easurements of {F}ractional {Q}uantum {H}all {E}nergy {G}aps in {E}dgeless {G}raphene {D}evices}.
\newblock \emph{\bibinfo{journal}{Phys. Rev. Lett.}} \textbf{\bibinfo{volume}{121}}, \bibinfo{pages}{226801} (\bibinfo{year}{2018}).

\bibitem{Kellogg2004}
\bibinfo{author}{Kellogg, M.}, \bibinfo{author}{Eisenstein, J.~P.}, \bibinfo{author}{Pfeiffer, L.~N.} \& \bibinfo{author}{West, K.~W.}
\newblock \bibinfo{title}{Vanishing {H}all {R}esistance at {H}igh {M}agnetic {F}ield in a {D}ouble-{L}ayer {T}wo-{D}imensional {E}lectron {S}ystem}.
\newblock \emph{\bibinfo{journal}{Phys. Rev. Lett.}} \textbf{\bibinfo{volume}{93}}, \bibinfo{pages}{036801} (\bibinfo{year}{2004}).

\bibitem{Tutuc2004counterflow}
\bibinfo{author}{Tutuc, E.}, \bibinfo{author}{Shayegan, M.} \& \bibinfo{author}{Huse, D.~A.}
\newblock \bibinfo{title}{Counterflow {M}easurements in {S}trongly {C}orrelated {G}a{A}s {H}ole {B}ilayers: {E}vidence for {E}lectron-{H}ole {P}airing}.
\newblock \emph{\bibinfo{journal}{Phys. Rev. Lett.}} \textbf{\bibinfo{volume}{93}}, \bibinfo{pages}{036802} (\bibinfo{year}{2004}).

\bibitem{Zeng2023solid}
\bibinfo{author}{Zeng, Y.} \emph{et~al.}
\newblock \bibinfo{title}{Evidence for a {S}uperfluid-to-solid {T}ransition of {B}ilayer {E}xcitons}.
\newblock \emph{\bibinfo{journal}{arXiv:2306.16995}}  (\bibinfo{year}{2023}).

\bibitem{Jain.03}
\bibinfo{author}{Jain, J.~K.}
\newblock \emph{\bibinfo{title}{Composite Fermions}} (\bibinfo{publisher}{Cambridge University Press}, \bibinfo{year}{2003}).

\bibitem{Jain2015CF}
\bibinfo{author}{Jain, J.~K.}
\newblock \bibinfo{title}{Composite fermion theory of exotic fractional quantum {H}all effect}.
\newblock \emph{\bibinfo{journal}{Annu. Rev. Condens. Matter Phys.}} \textbf{\bibinfo{volume}{6}}, \bibinfo{pages}{39--62} (\bibinfo{year}{2015}).

\bibitem{Eisenstein1990FQHE}
\bibinfo{author}{Eisenstein, J.} \& \bibinfo{author}{Stormer, H.}
\newblock \bibinfo{title}{The fractional quantum {H}all effect}.
\newblock \emph{\bibinfo{journal}{Science}} \textbf{\bibinfo{volume}{248}}, \bibinfo{pages}{1510--1516} (\bibinfo{year}{1990}).

\bibitem{Liu2019interlayer}
\bibinfo{author}{Liu, X.} \emph{et~al.}
\newblock \bibinfo{title}{Interlayer fractional quantum {H}all effect in a coupled graphene double layer}.
\newblock \emph{\bibinfo{journal}{Nat. Phys.}} \textbf{\bibinfo{volume}{15}}, \bibinfo{pages}{893--897} (\bibinfo{year}{2019}).

\bibitem{Scarola2001}
\bibinfo{author}{Scarola, V.~W.} \& \bibinfo{author}{Jain, J.~K.}
\newblock \bibinfo{title}{Phase diagram of bilayer composite fermion states}.
\newblock \emph{\bibinfo{journal}{Phys. Rev. B}} \textbf{\bibinfo{volume}{64}}, \bibinfo{pages}{085313} (\bibinfo{year}{2001}).

\bibitem{Huse2005}
\bibinfo{author}{Huse, D.~A.}
\newblock \bibinfo{title}{Resistance due to vortex motion in the $\nu$= 1 bilayer quantum {H}all superfluid}.
\newblock \emph{\bibinfo{journal}{Phys. Rev. B}} \textbf{\bibinfo{volume}{72}}, \bibinfo{pages}{064514} (\bibinfo{year}{2005}).

\bibitem{Kellogg2005}
\bibinfo{author}{Kellogg, M.~J.}
\newblock \emph{\bibinfo{title}{Evidence for {E}xcitonic {S}uperfluidity in a {B}ilayer {T}wo-{D}imensional {E}lectron {S}ystem}}.
\newblock Ph.D. thesis, \bibinfo{school}{California Institute of Technology} (\bibinfo{year}{2005}).

\bibitem{KT1973}
\bibinfo{author}{Kosterlitz, J.~M.} \& \bibinfo{author}{Thouless, D.~J.}
\newblock \bibinfo{title}{Ordering, metastability and phase transitions in two-dimensional systems}.
\newblock \emph{\bibinfo{journal}{J. Phys. C: Solid State Phys.}} \textbf{\bibinfo{volume}{6}}, \bibinfo{pages}{1181} (\bibinfo{year}{1973}).

\bibitem{Berezinskii1972BKT}
\bibinfo{author}{Berezinskii, V.}
\newblock \bibinfo{title}{Destruction of {L}ong-range {O}rder in {O}ne-dimensional and {T}wo-dimensional {S}ystems {P}ossessing a {C}ontinuous {S}ymmetry {G}roup. {II}. {Q}uantum {S}ystems}.
\newblock \emph{\bibinfo{journal}{Sov. Phys. JETP}} \textbf{\bibinfo{volume}{34}}, \bibinfo{pages}{610--616} (\bibinfo{year}{1972}).

\bibitem{Wen2004quantum}
\bibinfo{author}{Wen, X.-G.}
\newblock \emph{\bibinfo{title}{Quantum field theory of many-body systems: From the origin of sound to an origin of light and electrons}} (\bibinfo{publisher}{OUP Oxford}, \bibinfo{year}{2004}).

\bibitem{review-edge}
\bibinfo{author}{Heiblum, M.} \& \bibinfo{author}{Feldman, D.~E.}
\newblock \bibinfo{title}{Edge probes of topological order}.
\newblock \emph{\bibinfo{journal}{Int. J. Mod. Phys. A}} \textbf{\bibinfo{volume}{35}}, \bibinfo{pages}{2030009} (\bibinfo{year}{2020}).

\bibitem{Heiblum.10}
\bibinfo{author}{Bid, A.} \emph{et~al.}
\newblock \bibinfo{title}{Observation of neutral modes in the fractional quantum {H}all regime}.
\newblock \emph{\bibinfo{journal}{Nature}} \textbf{\bibinfo{volume}{466}}, \bibinfo{pages}{585--590} (\bibinfo{year}{2010}).

\bibitem{Takei2014Spinsuperfluid}
\bibinfo{author}{Takei, S.}, \bibinfo{author}{Halperin, B.~I.}, \bibinfo{author}{Yacoby, A.} \& \bibinfo{author}{Tserkovnyak, Y.}
\newblock \bibinfo{title}{Superfluid spin transport through antiferromagnetic insulators}.
\newblock \emph{\bibinfo{journal}{Phys. Rev. B}} \textbf{\bibinfo{volume}{90}}, \bibinfo{pages}{094408} (\bibinfo{year}{2014}).

\bibitem{Takei2016Spinsuperfluid}
\bibinfo{author}{Takei, S.}, \bibinfo{author}{Yacoby, A.}, \bibinfo{author}{Halperin, B.~I.} \& \bibinfo{author}{Tserkovnyak, Y.}
\newblock \bibinfo{title}{Spin {S}uperfluidity in the $\ensuremath{\nu}=0$ {Q}uantum {H}all {S}tate of {G}raphene}.
\newblock \emph{\bibinfo{journal}{Phys. Rev. Lett.}} \textbf{\bibinfo{volume}{116}}, \bibinfo{pages}{216801} (\bibinfo{year}{2016}).

\bibitem{Wei2018spinwave}
\bibinfo{author}{Wei, D.~S.} \emph{et~al.}
\newblock \bibinfo{title}{Electrical generation and detection of spin waves in a quantum {H}all ferromagnet}.
\newblock \emph{\bibinfo{journal}{Science}} \textbf{\bibinfo{volume}{362}}, \bibinfo{pages}{229--233} (\bibinfo{year}{2018}).

\bibitem{Zhou2022spinwave}
\bibinfo{author}{Zhou, H.} \emph{et~al.}
\newblock \bibinfo{title}{Strong-{M}agnetic-{F}ield {M}agnon {T}ransport in {M}onolayer {G}raphene}.
\newblock \emph{\bibinfo{journal}{Phys. Rev. X}} \textbf{\bibinfo{volume}{12}}, \bibinfo{pages}{021060} (\bibinfo{year}{2022}).

\bibitem{Cai2023}
\bibinfo{author}{Cai, J.} \emph{et~al.}
\newblock \bibinfo{title}{Signatures of fractional quantum anomalous {H}all states in twisted {M}o{T}e$_2$}.
\newblock \emph{\bibinfo{journal}{Nature}} \textbf{\bibinfo{volume}{622}}, \bibinfo{pages}{63--68} (\bibinfo{year}{2023}).

\bibitem{Park2023}
\bibinfo{author}{Park, H.} \emph{et~al.}
\newblock \bibinfo{title}{Observation of fractionally quantized anomalous {H}all effect}.
\newblock \emph{\bibinfo{journal}{Nature}} \textbf{\bibinfo{volume}{622}}, \bibinfo{pages}{74--79} (\bibinfo{year}{2023}).

\bibitem{Xu2023}
\bibinfo{author}{Xu, F.} \emph{et~al.}
\newblock \bibinfo{title}{Observation of {I}nteger and {F}ractional {Q}uantum {A}nomalous {H}all {E}ffects in {T}wisted {B}ilayer {M}o{T}e$_2$}.
\newblock \emph{\bibinfo{journal}{Phys. Rev. X}} \textbf{\bibinfo{volume}{13}}, \bibinfo{pages}{031037} (\bibinfo{year}{2023}).

\bibitem{Zeng2023}
\bibinfo{author}{Zeng, Y.} \emph{et~al.}
\newblock \bibinfo{title}{Thermodynamic evidence of fractional {C}hern insulator in moir{\'e} {M}o{T}e$_2$}.
\newblock \emph{\bibinfo{journal}{Nature}} \textbf{\bibinfo{volume}{622}}, \bibinfo{pages}{69--73} (\bibinfo{year}{2023}).

\bibitem{Lu2024}
\bibinfo{author}{Lu, Z.} \emph{et~al.}
\newblock \bibinfo{title}{Fractional quantum anomalous {H}all effect in multilayer graphene}.
\newblock \emph{\bibinfo{journal}{Nature}} \textbf{\bibinfo{volume}{626}}, \bibinfo{pages}{759--764} (\bibinfo{year}{2024}).

\bibitem{Eisenstein1989}
\bibinfo{author}{Eisenstein, J.~P.}, \bibinfo{author}{Stormer, H.~L.}, \bibinfo{author}{Pfeiffer, L.} \& \bibinfo{author}{West, K.~W.}
\newblock \bibinfo{title}{Evidence for a phase transition in the fractional quantum {H}all effect}.
\newblock \emph{\bibinfo{journal}{Phys. Rev. Lett.}} \textbf{\bibinfo{volume}{62}}, \bibinfo{pages}{1540--1543} (\bibinfo{year}{1989}).

\bibitem{Eisenstein1990spin}
\bibinfo{author}{Eisenstein, J.}, \bibinfo{author}{Stormer, H.}, \bibinfo{author}{Pfeiffer, L.} \& \bibinfo{author}{West, K.}
\newblock \bibinfo{title}{Evidence for a spin transition in the $\nu$= 2/3 fractional quantum {H}all effect}.
\newblock \emph{\bibinfo{journal}{Phys. Rev. B}} \textbf{\bibinfo{volume}{41}}, \bibinfo{pages}{7910} (\bibinfo{year}{1990}).

\bibitem{Engel1992}
\bibinfo{author}{Engel, L.~W.}, \bibinfo{author}{Hwang, S.~W.}, \bibinfo{author}{Sajoto, T.}, \bibinfo{author}{Tsui, D.~C.} \& \bibinfo{author}{Shayegan, M.}
\newblock \bibinfo{title}{Fractional quantum {H}all effect at \ensuremath{\nu}=2/3 and 3/5 in tilted magnetic fields}.
\newblock \emph{\bibinfo{journal}{Phys. Rev. B}} \textbf{\bibinfo{volume}{45}}, \bibinfo{pages}{3418--3425} (\bibinfo{year}{1992}).

\bibitem{Champagne.08b}
\bibinfo{author}{Champagne, A.~R.}, \bibinfo{author}{Finck, A. D.~K.}, \bibinfo{author}{Eisenstein, J.~P.}, \bibinfo{author}{Pfeiffer, L.~N.} \& \bibinfo{author}{West, K.~W.}
\newblock \bibinfo{title}{Charge imbalance and bilayer two-dimensional electron systems at ${\ensuremath{\nu}}_{T}=1$}.
\newblock \emph{\bibinfo{journal}{Phys. Rev. B}} \textbf{\bibinfo{volume}{78}}, \bibinfo{pages}{205310} (\bibinfo{year}{2008}).

\end{thebibliography}

\newpage
\clearpage

\section*{Method}

\renewcommand{\thefigure}{M\arabic{figure}}
\def\theequation{M\arabic{equation}}
\def\thetable{M\Roman{table}}
\setcounter{figure}{0}
\setcounter{equation}{0}

\subsection{0. Notations of LL filling}

The presence of layer pseudo-spin unlocks an extra dimension in the phase space of FQHE, which requires defining proper notation to describe the FQHE sequence. Here, we use the \CFi\ sequence as an example to explain the notation of LL filling and its connection with different FQHE sequences. 

\begin{figure*}
\includegraphics[width=0.9\linewidth]{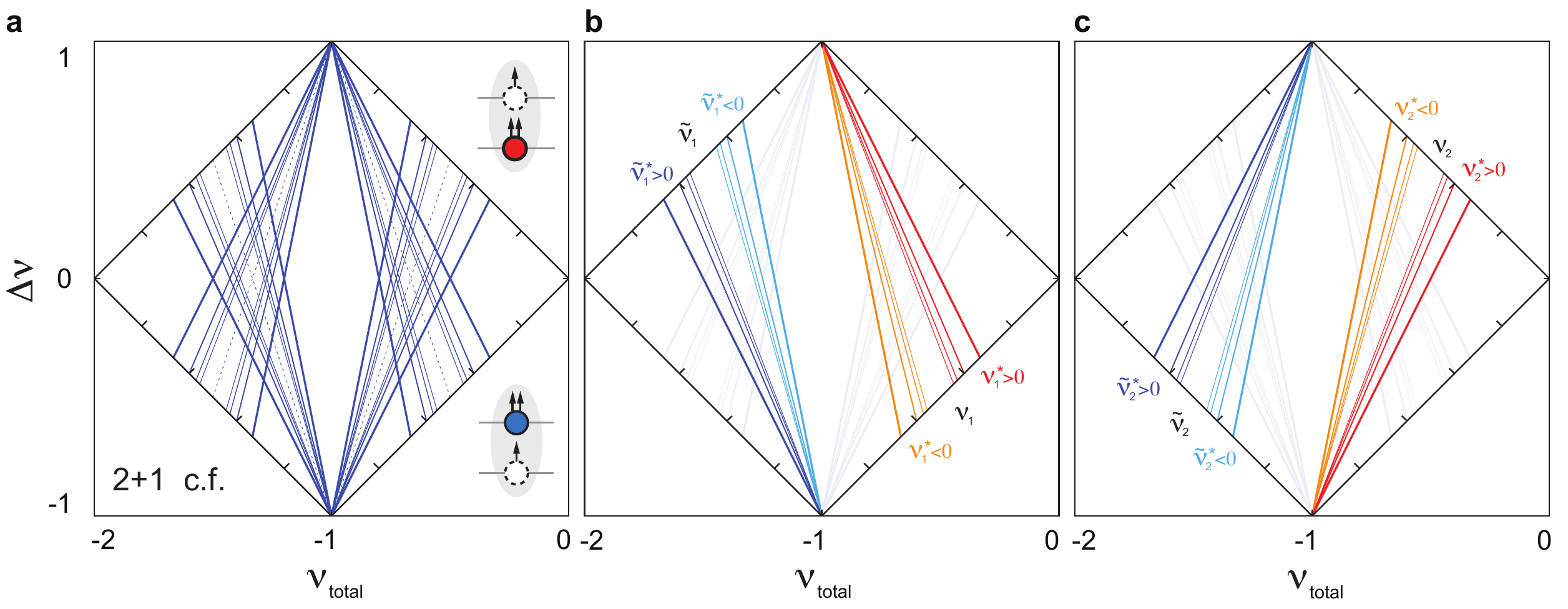}
\caption{\label{notation}{\bf{Notations of LL filling across the $\nu_{\text{total}}-\Delta\nu$ map.}} 
(a) Schematic map of \CFi\ states, which is the same as Fig.~\ref{fig1}f. (b-c) The same schematic map highlighting trajectories defined by integer values of (b) $^2_1\nu^{\ast}_1$ and (c) $^2_1\nu^{\ast}_2$. In panels (b) and (c), different colors highlight four distinct sequences in each layer. Red and orange denote the FQHE sequence of majority charge carriers, which are the hole-type carriers in the chosen filling fraction of $-2 < \nu_{\text{total}} < 0$. Blue and teal represent the FQHE sequence of minority charge carriers. Furthermore, blue and red mark the sequence defined by positive effective filling, whereas orange and teal label the sequence with negative effective filling.   }
\end{figure*}

Fig.~\ref{fig1}f plots the schematic diagram of FQHE states with the \CFi\ construction. This 2D map consists of $8$ separate FQHE sequences, which are marked in Fig.~\ref{notation}. The range of the 2D map is determined by the first integer filling of hole-doped LL. In the range of  $-1 < \nu_{\text{total}} < 0$, FQHE states can be viewed as effective integer quantum Hall effect (IQHE) of hole-type CF carriers, following the definition of Eq.~(\ref{Eq:CF}). In this range, four FQHE sequences are defined by ${}^2_1\nu^{\ast}_1 >0$ (red traces in Fig.~\ref{notation}b), ${}^2_1\nu^{\ast}_1 <0$ (orange traces in Fig.~\ref{notation}b), ${}^2_1\nu^{\ast}_2 >0$ (red traces in Fig.~\ref{notation}c), ${}^2_1\nu^{\ast}_2 <0$ (orange traces in Fig.~\ref{notation}c). 

On the other hand, FQHE states in the range of  $-2 < \nu_{\text{total}} < -1$ does not comply with the definition of Eq.~(\ref{Eq:CF}) directly. These FQHE states can be understood by considering the particle-hole conjugate of Eq.~(\ref{Eq:CF}). Instead of carrier filling $\nu$, we consider the LL filling of minority carrier $\tilde{\nu}$, defined as $\tilde{\nu}_1=1+\nu_1,$ and $\tilde{\nu}_2=1+\nu_2$. Following this definition, Eq.~(\ref{Eq:CF}) is rewritten as,
\begin{equation}\label{Eq:CF_tilde}
    {}^a_b\tilde{\nu}_1^{\ast} = \frac{\tilde{\nu}_1}{1-a\tilde{\nu}_1-b\tilde{\nu}_2}, \ \ {}^a_b\tilde{\nu}_2^{\ast} = \frac{\tilde{\nu}_2}{1-a\tilde{\nu}_2-b\tilde{\nu}_1}.
\end{equation}
FQHE states in the range of $-2 < \nu_{\text{total}} < -1$, therefore, emerge along trajectories defined by ${}^a_b\tilde{\nu}_1^{\ast}$ and ${}^a_b\tilde{\nu}_2^{\ast}$ taking positive and negative integer values. 

According to Fig.~\ref{fig1}d, the FQHE sequences defined by Eq.~(\ref{Eq:CF}) ($-1 < \nu_{\text{total}} < 0$) and Eq.~(\ref{Eq:CF_tilde}) ($-2 < \nu_{\text{total}} < -1$) exhibit excellent symmetry around \nutot $=-1$. Such symmetry is also noted in previous observations of 2-component FQHE in quantum Hall graphene bilayer ~\cite{Li2019pairing}. 

Given the symmetry around \nutot $=-1$, this work focuses on examining the FQHE in the range of $-2 < \nu_{\text{total}} < -1$. This choice is also motivated by an experimental constraint: electric contacts to the graphene bilayer tend to be more consistent in  the range of $-2 < \nu_{\text{total}} < -1$. This constraint is especially prominent in counterflow and drag geometries. 

Following the definition of $\tilde{\nu}_1$ and $\tilde{\nu}_2$, we have $\tilde{\nu}_{\text{total}}=2+\nu_{\text{total}}$ and $\Delta\nu = \nu_1 - \nu_2 = \tilde{\nu}_1 - \tilde{\nu}_2$. The transport responses shown in Fig.~\ref{fig4}b is measured along a trajectory marked by the blue dashed line in Fig.~\ref{notation}c. Tracing constant value of $^2_1\tilde{\nu}_1^{\ast} = 1$, this trajectory goes through two FQHE sequences defined by $^2_1\tilde{\nu}_2^{\ast} >0$ (dark blue lines in Fig.~\ref{notation}c) and $^2_1\tilde{\nu}_2^{\ast} <0$ (light blue lines in Fig.~\ref{notation}c). 

The same sequences are observed along another trajectory defined by $^2_1\tilde{\nu}_1^{\ast} = 2$ (see Fig.~\ref{CF21}). Due to the interlayer flux attachment in the \CFi\ construction, changing $^2_1\tilde{\nu}_1^{\ast}$ from $1$ to $2$ generates a shift in the location of FQHE states defined by $^2_1\tilde{\nu}_2^{\ast} = N$. This shift is demonstrated by Fig.~\ref{CF21}b and c.

\begin{figure*}
\includegraphics[width=1\linewidth]{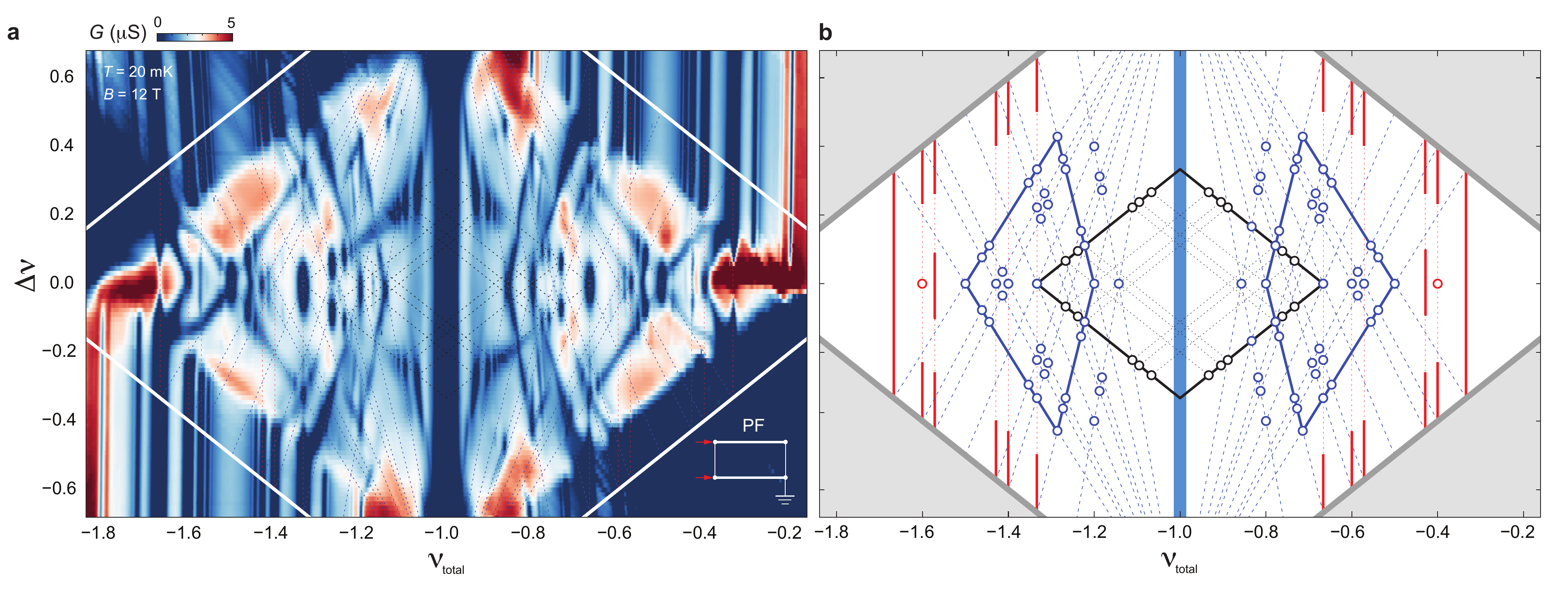}
\caption{\label{CFmap}{\bf{Two-component FQHE states with different CF constructions. }} (a) $G_{\textrm{PF}}$ as a function of \nutot\ and \dnu\ at $B = 12$T and $T = 20 $mK. (b) Schematic diagram showing FQHE associated with the \CFn (black), \CFi (blue), and \CFii (red) constructions. Grey solid lines mark IQHE states at constant integer  LL fillings. Together, these IQHE states define the diamond shaped phase space that is of interests in this work. All two-component FQHE states emerge within this phase space. The excitonic state described by the $(111)$ wavefunction is observed along the vertical trajectory at $\nu_{\text{total}} = -1$, which is marked by the blue vertical stripe in (b). Effective IQHE of \CFi\ and \CFn\ are marked by blue and black dashed lines, which correspond to constant integer values of ${}^2_1\nu_{i}$ and ${}^2_0\nu_{i}$. The effective excitonic pairing between \CFii leads to FQHE states along red dashed lines, which are defined by constant integer value of ${}^2_2\nu^{\ast}_{1}+{}^2_2\nu^{\ast}_{2}={}^2_2\nu^{\ast}_{\text{total}}$. By matching expected trajectories of different CF constructions, we are able to identify the origin of FQHE states and the composition of fractional excitons.  }
\end{figure*}

\begin{figure*}
\includegraphics[width=0.57\linewidth]{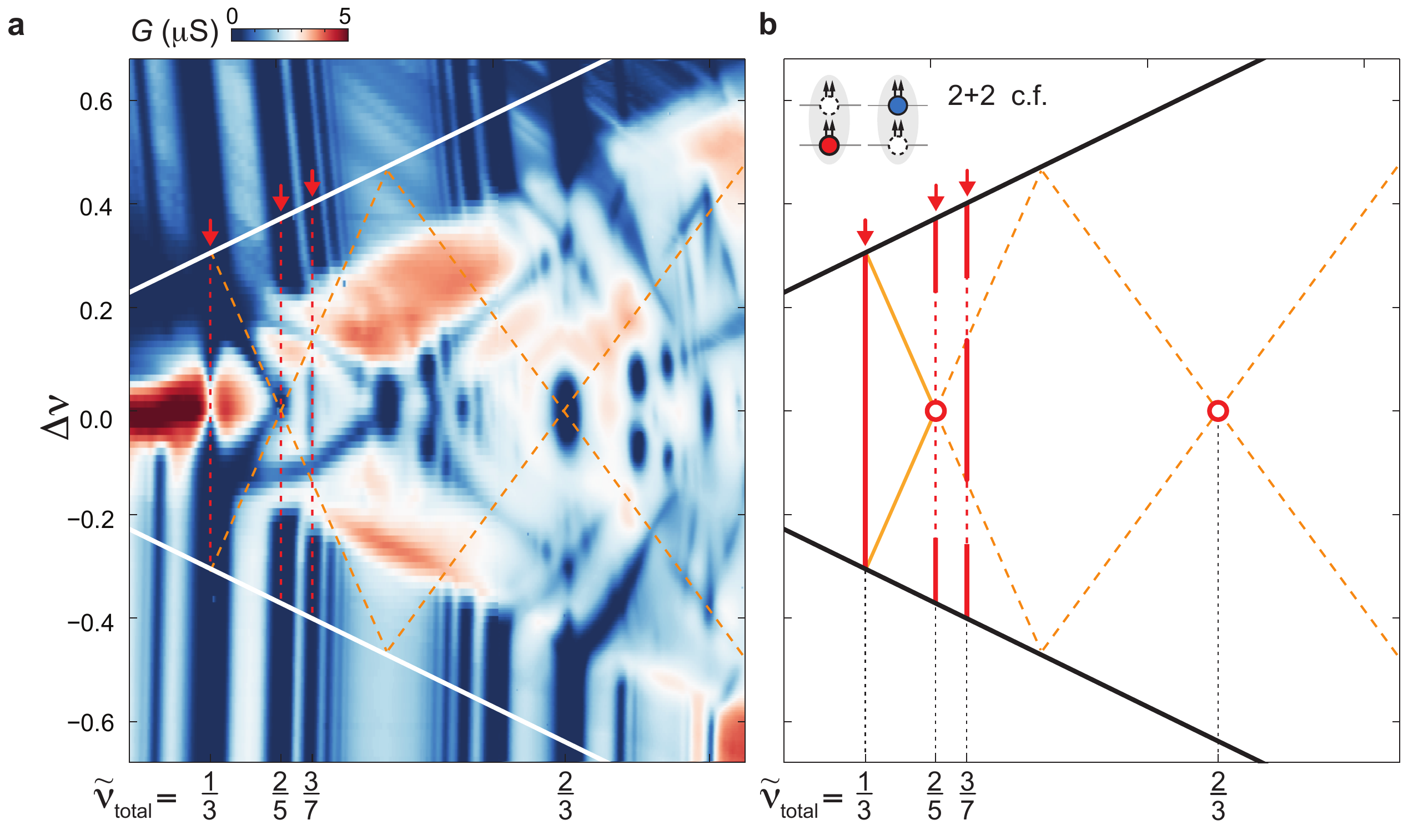}
\caption{\label{CF22}{\bf{Effective IQHE of \CFii\ and exciton pairing between \CFii. }} (a) $G_{\textrm{PF}}$ as a function of \nutot\ and \dnu\ at $B = 12$T and $T = 20 $mK. (b) Schematic diagram marks the most prominent features associated with the \CFii constructions. Orange dashed lines correspond to constant integer values of ${}^2_2\nu_{1}^{\ast}$ and ${}^2_2\nu^{\ast}_{2}$, which mark the expected trajectory of effective IQHE of \CFii. This is distinct from the \CFii\ excitons. When the sum of effective $\Lambda$-level fillings across two graphene layers equals an integer value, an emerging FQHE state is directly comparable with the exciton state at total filling of one. At ${}^2_2\nu^{\ast}_\textrm{total} = N \in \mathbb{Z}$, the ground state order is described by exciton pairing between \CFii. At $\tilde{\nu}_\textrm{total}=2/5$, we observe transitions between \CFii\ exciton, marked by vertical red solid lines, and an effective IQHE of \CFii, marked by red open circle in (b). Along the same vein, a series of transitions are observed at $\tilde{\nu}_\textrm{total}=3/7$. Notably, \CFi\ and \CFn\ constructions are unlikely to produce incompressible states in the portion of the phase space near $\tilde{\nu}_\textrm{total}=2/5$. Therefore, the observation of FQHE states along the expected trajectories of \CFii\ states provides strong support for the \CFii\ construction.   }
\end{figure*}

\begin{figure*}
\includegraphics[width=0.85\linewidth]{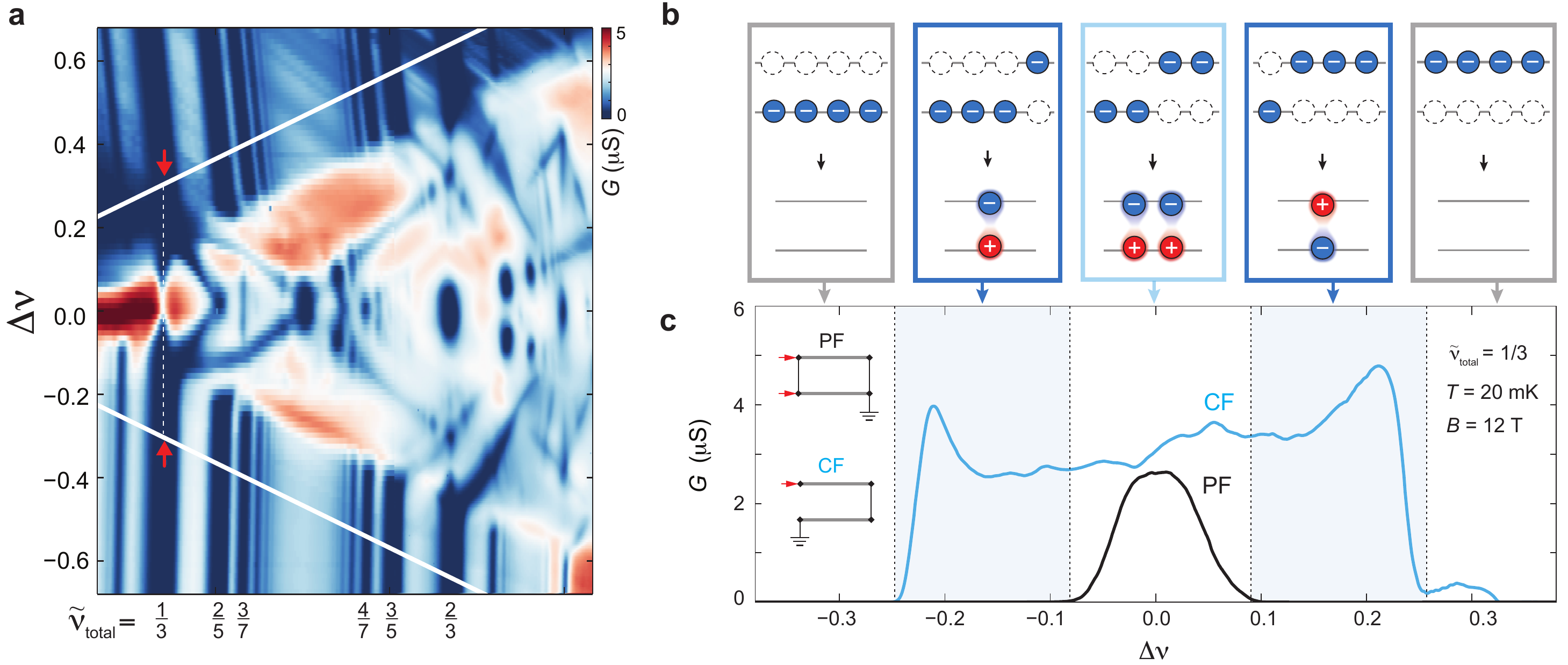}
\caption{\label{fig333}{\bf{$(333)$ wavefunction and the fractional analogue of the $(111)$ exciton condensate.}} (a) Schematic diagram of the (333) wavefunction at $\tilde{\nu}_1 + \tilde{\nu}_2 = 1/3$ (top), which is constructed from a single-layer FQHE state (middle) combined with interlayer excitons (bottom). (b) According to the construction in (a), the FQHE at $\tilde{\nu}_{\text{total}}  = 1/3$ exhibits different exciton density while maintaining the same excitonic order across different \dnu-regimes. (c) $G_{\text{PF}}$ (black trace) and  $G_{\text{CF}}$ (blue trace) as a function of \dnu\ measured at $\tilde{\nu}_{\text{total}} = 1/3$. Blue shaded background marks regimes where robust charge gap coexists with charge-neutral exciton flow. (d-e) $I_{\text{drive}}$ and $I_{\text{drag}}$ as a function of d.c. voltage bias measured (d) for the excitonic state at \dnu $=-0.2$ and (e) in the single-layer regime at \dnu $=-0.33$. (f) $G_{\text{PF}}$ (top panel), drag ratio $I_{\text{drag}}/I_{\text{drive}}$ (middle panel), and $G_{\text{CF}}$ (bottom panel) as a function of $T$ measured at $\tilde{\nu}_{\text{total}} = 1/3$ and \dnu $= -0.20$. }
\end{figure*}

\subsection{I. Two types of CF states}

FQHE states reported here fall into two categories: (i) effective IQHE of CFs  emerges from fully occupied $\Lambda$-level in each graphene layer; (ii) effective excitonic pairing between CFs arises when the sum of effective $\Lambda$-level filling across two graphene layers equals an integer value. This is directly comparable to the IQHE, when the LL in a single layer is fully occupied (grey solid lines in Fig.~\ref{CFmap}b), and the exciton state, when the sum of LL filling across two graphene layers equal to one (vertical blue stripe in Fig.~\ref{CFmap}b). 

Effective IQHE of \CFi\ and \CFn\ are marked by blue and black dashed lines in Fig.~\ref{CFmap}b, which correspond to constant integer values of ${}^2_1\nu_{i}^{\ast}$ and ${}^2_0\nu_{i}^{\ast}$. Along blue (black) dashed lines, \CFi\ (\CFn) fully occupy an integer number of $\Lambda$-levels. When integer ${}^2_1\nu_{i}^{\ast}$ is realized in only one graphene layer, one graphene layer is incompressible while the other is compressible. In this scenario, the quantum Hall bilayer exhibits a phenomenon called semi-quantization ~\cite{Liu2019interlayer}. When integer ${}^2_1\nu_{i}^{\ast}$ is realized in both graphene layers, a robust charge gap develops across both layers of the quantum Hall bilayer structure, which gives rise to insulating features with vanishing $G_{\text{PF}}$ and are marked by open circles in Fig.~\ref{CFmap}b. A sequence of these insulating features form the 2D pattern of FQHE states in the \nutot-\dnu\ map (Fig.~\ref{CFmap}a-b). 

Effective excitonic pairing is observed between \CFii. Exciton pairing between \CFii\ gives rise to FQHE states along red dashed lines in Fig.~\ref{CF22}a, which are defined by constant integer value of ${}^2_2\nu_{1}^{\ast}+{}^2_2\nu_{2}^{\ast}={}^2_2\nu^{\ast}_{\text{total}}$. In contrast, constant integer values of ${}^2_2\nu_{i}^{\ast}$ are marked by orange dashed lines in Fig.~\ref{CF22}. The distinction between orange and red dashed lines offers unambiguous evidence for excitonic pairing between composite particles and holes following the \CFii\ construction. The \CFii\ excitons  are fractional analogue of the (111) state and can be directly compared with the excitonic state along the vertical trajectory at total filling of one. 

Fig.~\ref{fig333} examines the FQHE at $\tilde{\nu}_{\text{total}}= 1/3$ as a function of varying \dnu. We show that this 2-component FQHE is in excellent agreement with the Halperin (333) wavefunction. According to previous theoretical discussion, this FQHE is expected to emerge at $\tilde{\nu}_1 + \tilde{\nu}_2 = 1/3$ ~\cite{Wen1992}. Following the construction in Fig.~\ref{fig2}a, a (333) state (top panel in Fig.~\ref{fig2}a) can be viewed as a combination of the Laughlin state in one layer (middle panel in Fig.~\ref{fig2}a), and an ensemble of interlayer excitons (bottom panel in Fig.~\ref{fig2}a). According to this construction, varying \dnu\ only changes the excitonic density, while having no impact on the excitonic nature of the FQHE state.  Consequently, the (333) state is expected to be stable over a wide range of \dnu. 

Fig.~\ref{fig333}a plots the PF conductance $G_{\text{PF}}$ as a function of \dnu\ and $\tilde{\nu}_{\text{total}}$. In the two-component regime, the  FQHE state at $\tilde{\nu}_{\text{total}}= 1/3$, indicated by the red arrows and vertical dashed line, is shown to extend along constant total filling, occupying the entire \dnu\ range. This robustness against \dnu\ is in alignment with the expected $(333)$ state.  

The nature of the ground state is revealed by transport responses measured with varying \dnu. A robust excitonic state is observed in the \dnu\ regime marked by the blue shaded background (Fig.~\ref{fig333}b), where vanishing $G_{\text{PF}}$ (black trace) coincides with non-zero counterflow conductance $G_{\text{CF}}$ (blue trace). This background underscores the stability of \CFii\ excitons against layer imbalance. 

Near \dnu $=0$, a reduction in the charge gap gives rise to free charge carriers, resulting in non-zero $G_{\text{PF}}$ and a reduced drag response. Despite this reduction in the charge gap, the ground state order remains unchanged across the entire \dnu\ range, as illustrated in Fig.~\ref{1over3}. It is important to note that the reduction in the charge gap near \dnu $=0$ is consistent with the $(111)$ state ~\cite{Li2017superfluid,Zeng2023solid}. The dependence on \dnu\ offers an indirect evidence supporting the $(333)$ wavefunction as the origin of the FQHE at $\tilde{\nu}_{\text{total}}$.

It should be emphasized that the excitonic order at \CFi\ states is distinct from \CFii\ excitons. 
For \CFii excitons, the appearance of a charge gap is directly related to the exciton pairing. For states \CFi, the emergence of a charge gap originates from the formation of $\Lambda$-levels. Indeed, the observation of excitonic pairing along \CFi\ states in Fig.~4 raises an interesting open question. In this paper, we discuss one of the most natural explanations by invoking exciton pairing between quasiparticle and quasihole excitations. This gives rise to excitons with fractionally charged constituents. 


\subsection{II. Measurement Configurations and Sample Geometry}

In this work, quantum Hall graphene bilayer samples are shaped into the edgeless Corbino geometry, as shown in Fig.~\ref{fig1}a. This geometry allows three distinct measurement geometry: parallel flow (PF), counterflow, and drag. The PF geometry probes the charge gap of the FQHE state. The presence of a robust FQHE state will be manifested in an insulating feature in PF geometry with vanishing bulk conductance $G_{\text{PF}}=0$. The counterflow and drag geometries probe excitonic flow in the charge-neutral counterflow channel. In both geometries, current flows in opposite directions across the top and bottom layers. For simplicity, we refer to the top layer as layer 1, and the bottom layer as layer 2.

\begin{figure}
\includegraphics[width=1\linewidth]{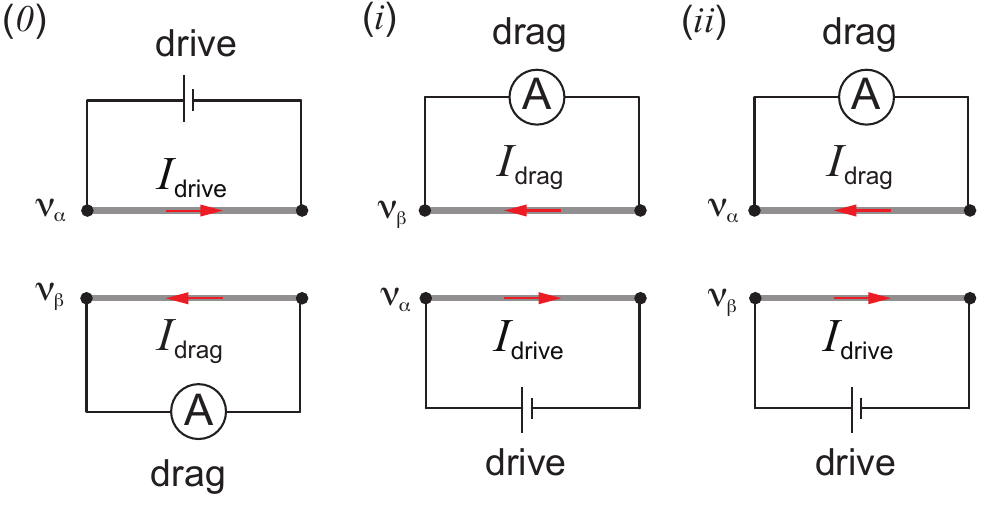}
\caption{\label{dragconfig}{\bf{Reversing drive and drag layers in the drag geometry. }} Three drag configurations are shown here. (0) represents the configuration used in the main text, where top graphene layer (layer 1) is the drive layer at LL filling $\nu_{\alpha}$, and bottom graphene layer (layer 2) is the drag layer at $\nu_{\beta}$. (i) is the first reversed configuration, where bottom graphene layer is the drive layer at $\nu_{\alpha}$ and top graphene layer is the drag layer at $\nu_{\beta}$. (ii) is the second reversed configuration,  bottom graphene layer is the drive layer at $\nu_{\beta}$ and top graphene layer is the drag layer at $\nu_{\alpha}$.  }
\end{figure}

\begin{figure*}
\includegraphics[width=1\linewidth]{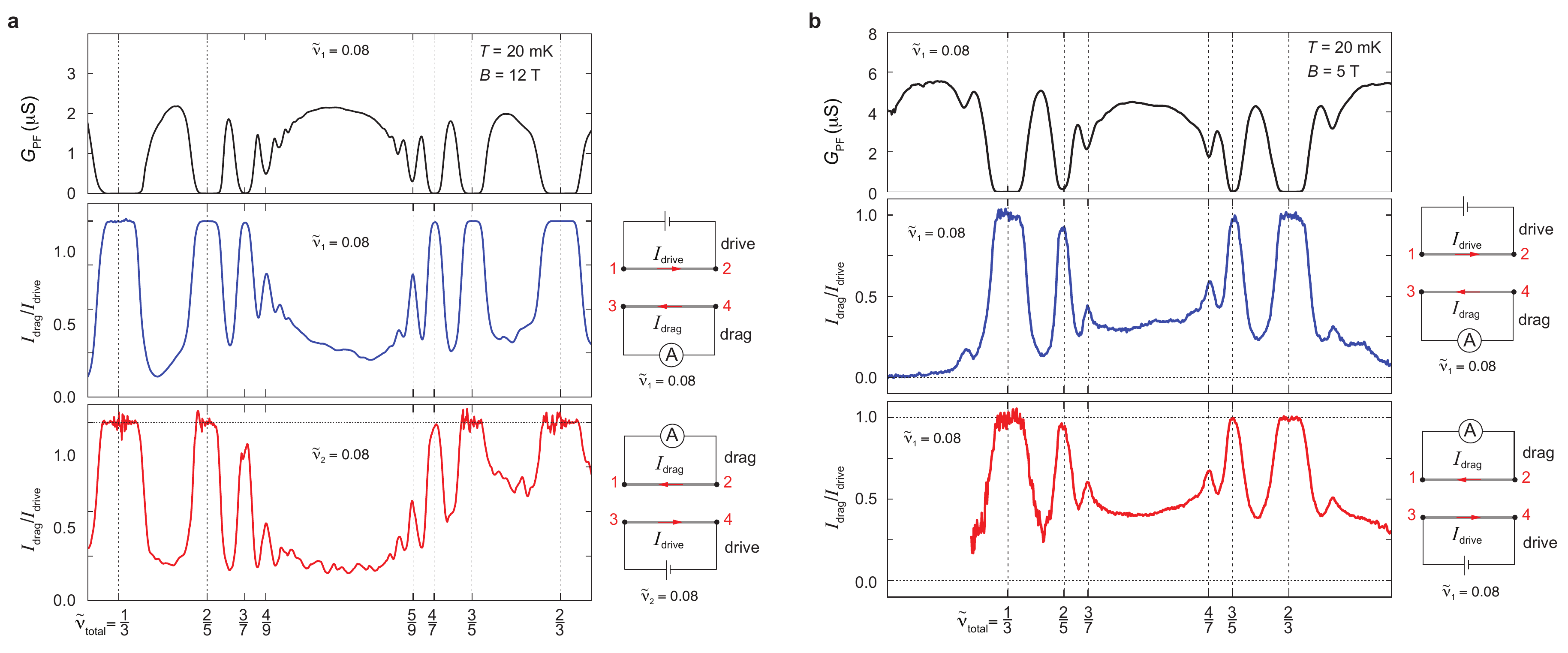}
\caption{\label{drivedrag}{\bf{Reversing drive and drag layers. }} (a) $G_{\text{PF}}$ (top), drag ratio $I_{\text{drag}}/I_{\text{drive}}$ measured from configuration (0) (middle), and drag ratio measured from configuration (i) (bottom) as a function of $\tilde{\nu}_{\text{total}}$ along the dashed orange line in Fig.~\ref{fig3}a.  (b) $G_{\text{PF}}$ (top), drag ratio $I_{\text{drag}}/I_{\text{drive}}$ measured from configuration (0) (middle), and drag ratio measured from configuration (ii) (bottom) as a function of $\tilde{\nu}_{\text{total}}$. (b) is measured from $B = 5$ T along the trajectory that is equivalent to the orange dashed line in Fig.~\ref{fig3}a. }
\end{figure*}

We discuss expected transport properties of a FQHE state, where a robust charge gap coexists with excitons that are overall charge neutral mode. Due to the robust charge gap, we expect to observe an insulating state in the PF geometry. The presence of interlayer excitons allows the counterflow measurement to directly couple to exciton flow, which generates a conductive response in the counterflow geometry. Most importantly, the coexistence of a robust charge gap and charge-neutral excitons will give rise to a characteristic transport behavior. Since current is carried solely by the counterflow mode of exciton flow, a current across the drive layer is perfectly converted into a current with equal amplitude and opposite direction across the drag layer (Fig.~\ref{dragconfig}). This unique transport response, described by $I_{\text{drag}}/I_{\text{drive}} = 1$, is referred to as the perfect drag response. If unpaired free charge is generated, the drag ratio will deviate from one, with $I_{\text{drag}}$ being smaller than $I_{\text{drive}}$.

Fig.~\ref{dragconfig} shows three drag configurations. Configuration (0) is utilized as the drag geometry in the main text, where the top graphene layer (layer 1) is the drive layer and the bottom graphene layer (layer 2) the drag layer. Along the orange line in Fig.~\ref{fig3}b and the blue line in Fig.~\ref{fig4}d, drive and drag layers are tuned to different LL fillings. Given a non-zero layer imbalance, there are two options for reversing the drive and drag layers. For simplicity, we denote the drive and drag layer filling in configuration (0) as $\nu_{\alpha}$ and  $\nu_{\beta}$. In configuration (i), bottom graphene layer is the drive layer at $\nu_{\alpha}$ and top graphene layer is the drag layer at $\nu_{\beta}$. Configuration (i) maintains the same LL fillings of drive and drag layers compared to configuration (0). In configuration (ii),  bottom graphene layer is the drive layer at $\nu_{\beta}$ and top graphene layer is the drag layer at $\nu_{\alpha}$. Configuration (ii) maintains the same LL fillings of the top and bottom graphene layers compared to configuration (0). 

As can be seen in Fig.~\ref{drivedrag}, the perfect drag response remains the same between three drag configurations, (0), (i), and (ii). Perfect drag ratio of one, $I_{\text{drag}}/I_{\text{drive}} = 1$, always coincides with the robust charge gap evidenced by vanishing $G_{\text{PF}}$. 

For the quantum Hall bilayer sample used in this work, interlayer separation is $4.5$ nm for the device with Corbino geometry.  The sample consists of graphite gate electrodes as the top most and bottom most encapsulating layer. 
Charge carrier density in layers 1 and 2, $n_{\text{1}}$ and $n_{\text{2}}$, can be independently controlled by applying voltage bias to top and bottom graphite gate electrodes. LL filling in each layer is defined as $\nu_i = n_i\Phi_0/B$. Here, $\Phi_0$ denotes the quantum of magnetic flux, $B$ is the external magnetic field. As such, $B/\Phi_0$ defines the number of magnetic flux penetrating a unit area of the sample.  

\subsection{III. Layer Pseudo-spin and Spin}


\begin{figure*}
\includegraphics[width=0.7\linewidth]{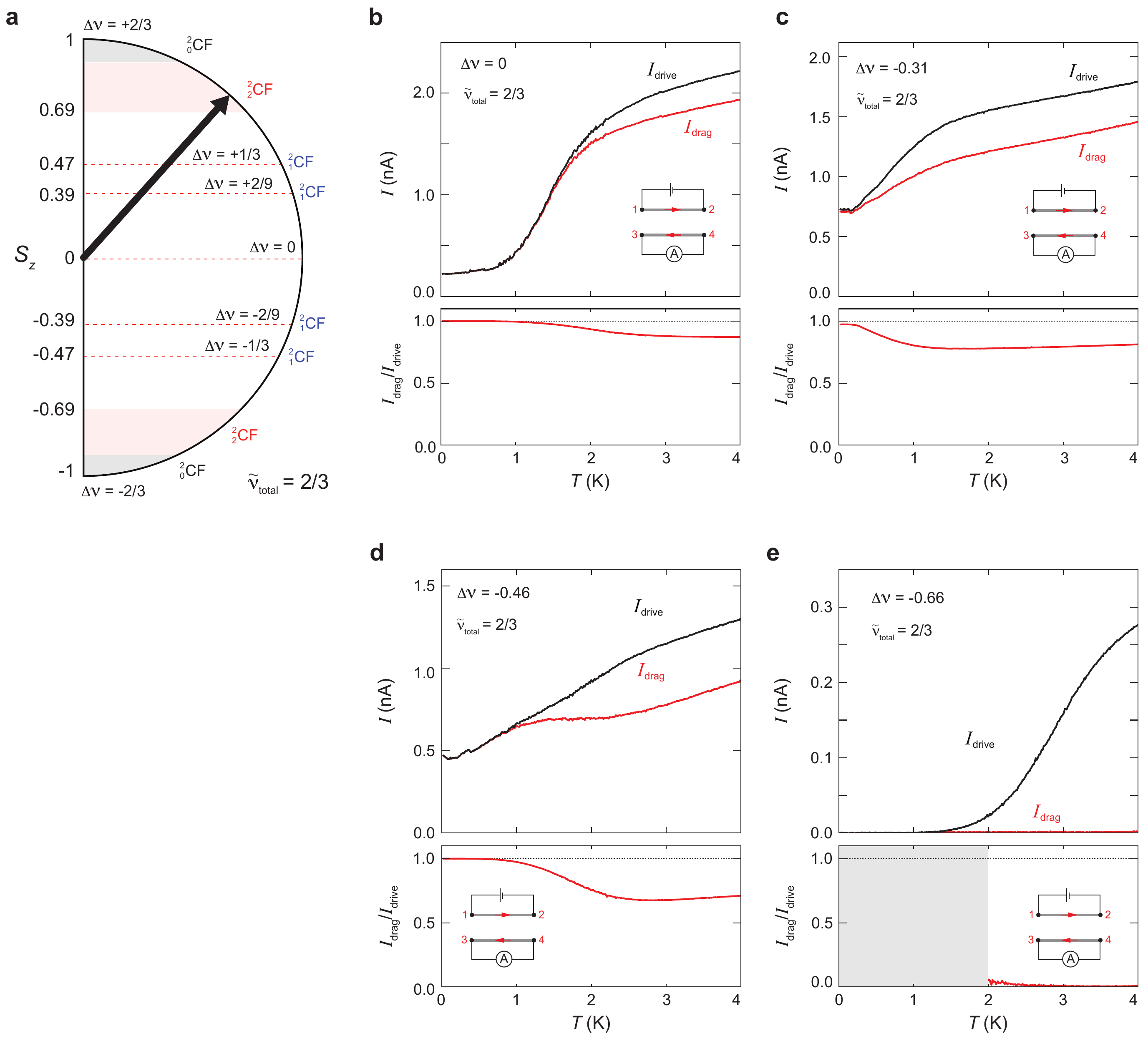}
\caption{\label{Sz}{\bf{Pseudo-spin transitions. }} (a) The evolution of FQHE orders with varying \dnu\ and $S_z$ at $\tilde{\nu}_{\text{total}} = 2/3$.  (b-e) $I_{\text{drive}}$ and $I_{\text{drag}}$ (top) and drag ratio $I_{\text{drag}}/I_{\text{drive}}$ (bottom) as a function of $T$ measured at  $\tilde{\nu}_{\text{total}} = 2/3$ and (b) \dnu $=0$, (c) \dnu $=-0.31$,  (d) \dnu $=-0.46$, and (e) \dnu $=-0.66$.  }
\end{figure*}

Given the analogy between pseudo-spin and spin, the insulating phase at \dnu $=0$ can be viwed as an antiferromagnetic insulator in the pseudospin channel. The presence of perfect drag response is indicative of a charge-neutral mode enabled by pairing between two pseudo-spin species. Since the energy gap at $\nu=0$ is close to $20$ K, quasiparticle excitation is negligible at $T = 20$ mK. As such, we argue that interlayer exciton is a gapless neutral mode. This enables an exciton condensate as the ground state. This condensate is the pseudo-spin equivalent of the spin superfluid, which is proposed to arise from charge neutral modes of an antiferromagnetic insulator ~\cite{Takei2014Spinsuperfluid,Takei2016Spinsuperfluid,Wei2018spinwave,Zhou2022spinwave}. Interestingly, transport measurements in quantum Hall bilayer uncover many excitonic states when $S_z$ deviates from $0$. Adopting the same language of spinful electrons, these excitonic states at non-zero \dnu\ correspond to canted-antiferromagnetic orders. 

Given the different dimensions between the phase space of spin and pseudo-spin, the analogy between pseudo-spin and spin is not intended to be exact. Nevertheless, our observations establish an unambiguous indication for a hidden dimension underlying the landscape of FQHE, which cannot be accessed by charge transport alone. While it remains an experimental challenge to directly probe the spin and valley order, or control their polarization, quantum Hall bilayer provides an ideal platform to investigate the nature of two-component FQHE and their neutral modes. 

At $\tilde{\nu}_{\text{total}} = 2/3$, counterflow and drag measurements reveals a series of psuedo-spin-driven transitions between FQHE orders with different exciton compositions. Different FQHE orders are marked by vertical arrows in Fig.~\ref{fig4}e. According to the \dnu-dependence of $G_{\text{PF}}$, the charge gap vanishes during the transition between \CFi\ and \CFii\ states. 

Interestingly, the \dnu-dependence at $\tilde{\nu}_{\text{total}} = 1/3$ is drastically different. At this filling, the charge gap remains robust across the transition from a single-layer Laughlin state to the excitonic order of \CFii (Fig.~\ref{fig2}c).

The dependence of 2-component FQHE on $S_z$ can be compared with the stability of single-layer FQHE against an inplane magnetic field ~\cite{Eisenstein1989,Eisenstein1990spin,Engel1992}. With increasing in-plane magnetic field, it was shown that the energy gap at $\nu=2/3$ diminishes before the emergence of another incompressible state, whereas the energy gap at $\nu=1/3$ appears unaffected by the in-plane $B$-field. In both cases, the application of an inplane field favors a spin-polarized state. Our observations in quantum Hall bilayer suggest that there are more viable configurations at $\nu=2/3$ compared to $\nu=1/3$. This creates an experimental basis for understanding the dependence on an in-plane magnetic field.

\subsection{IV. Theoretical Methods}

We consider Abelian states of the Halperin $(nnm) $ type and their generalizations. Here $n$ is an odd number, while $m$ is any integer  such that $m\le n$ for the state to be stable. The state is comprised of composite fermions, which carry $n-1$ quanta of intralayer magnetic flux and $m$ quanta of inter-layer flux.

In terms of the electron positions $x_k+iy_k$, the wave function is a product of an exponential factor, which is the same for all states, and  a polynomial, which depends on the topological order. To construct the polynomial, we divide electrons into two or more groups, so that the two layers correspond to different groups. If the positions of the electrons in the two groups are $z_a=x_a+iy_a$ and $w_b=x_b+iy_b$, the polynomial part of the $(nnm)$ wave function is $\Pi_{a>c}(z_a-z_c)^n\Pi_{b>d}(w_b-w_d)^n\Pi_{a,b}(z_a-w_b)^m$. The structure of the wave function is encoded in its ${\bf K}$-matrix \cite{Wen2004quantum}. 

\subsubsection{$(nnn)$ states}

An $(nnn)$ state has a degenerate ${\bf K}$-matrix, and its physics is different from other Halperin states. It describes a Laughlin state at the filling factor $1/n$, where electrons are allowed to reside in both layers. Such states are encountered for some combinations of the filling factors in this paper. Their neutral excitations are generated by moving composite fermions between two layers locally. Such excitations are bosons since they can be created by a local operation. Hence, we can think of them as excitons. Their charges in the two layers are simply equal to the charge of a composite fermion $\pm e$.

\subsubsection{Drag current}

For FQHE states following the \CFi\ construction, drag current can be generated either as a current of bilayer excitons or a backflow current due to interlayer flux attachment. 
Bilayer excitons combine opposite charges in the two layers. Thus, any motion of excitons results in the opposite electric currents in the two layers. This can naturally explain perfect drag. Alternatively, a drag current can emerge due to inter-layer flux attached to composite fermions. Thus, when composite fermions in one layer move, their attached flux moves, and its motion creates an effective electric field in the other layer. This field excites current in the second layer. 
We will see below that in the latter case, the drag current is less than the drive current, {\it i.e.}, the drag is not perfect.

As an example, we compute the backflow current in a system populated by composite fermions of positive charge $e$ carrying $2$ intralayer flux quanta and $1$ interlayer flux quantum. The composite fermion filling factors in the upper and lower layers are $\nu_1^*$ and $\nu_2^*$ respectively.
Consider the drive current density $j_{\rm drive}$ through the upper layer. 
Each moving composite fermion drags the attached magnetic flux with it. This induces an effective electric field.
To compute the drag current density $j_{\rm drag}$ in the lower layer, we find the effective electric field $E_{\text{eff}}$ generated by the currents $j_{\rm drive}$ and $j_{\rm drag}$ in the lower layer,
\begin{align}
E_{\text{eff}}=-(j_{\rm drive}+2j_{\rm drag})\frac{\Phi_0}{ec},
\end{align}
where $\Phi_0=hc/e$ is a flux quantum.
From the quantized conductance of the composite fermions,
\begin{align}\label{Eq:drag}
j_{\rm drag}=\frac{\nu_2^*e^2}{h}E_{\text{eff}}=-\frac{\nu_2^*}{1+2\nu_2^*}j_{\rm drive},
\end{align}
where we assumed zero $\sigma_{xx}$ in the lower layer.
Thus, the backflow current cannot provide perfect drag, which we ascribe to excitons. In the Supplemental text, we repeat the calculation for a small but non-zero longitudinal conductivity $\sigma_{xx}$ in both layers.

\subsubsection{Observed states in the bilayer system}

The $(nnm)$-type states, and their generalizations, with well-defined invertible ${\bf K}$-matrices are labeled in terms of the composite fermion $^a_b$CF filled levels. As is relevant to the observed data, we focus on the states belonging to $^2_0$CF, $^2_1$CF, and $^2_2$CF families. The ${\bf K}$-matrices, and their corresponding ${\bf t}$-vectors, for the bilayer states are described in terms of blocks, 
\begin{align}\label{Eq:Kmat}
    {\bf K} = \begin{pmatrix}
        {\bf K_1} & b {\bf J} \\
        b {\bf J}^T & {\bf K_2}
    \end{pmatrix},~~~~~~\text{and}~~~~~~{\bf t}={\bf t_1}\oplus {\bf t_2}.
\end{align}
The blocks ${\bf K_1}$ and ${\bf K_2}$, along with their respective charge vectors ${\bf t_1}$ and ${\bf t_2}$, describe the CF filled $\nu^*_{1}\in \mathbb{Z}$ and $\nu^*_{2}\in \mathbb{Z}$ monolayer states respectively. The operation $\oplus$ denotes direct sum. The matrix ${\bf J}$ is a constant matrix of ones with dimensions $\dim {\bf t_1}\times \dim {\bf t_2}$, and $b=0,1,2$. We work in the symmetric representation in which the ${\bf t}$ vector is ${\bf t}^T=(1,\dots,1)$ of dimension $\dim {\bf t_1}+\dim {\bf t_2}$. In the above construction, the total filling fraction is given by $\tilde \nu_{\text{t}} = {\bf t}^T {\bf K}^{-1} {\bf t}$ and the difference of the fillings is given by $\Delta \nu = {\bf t}^T {\bf K}^{-1} ({\bf t_1}\oplus {\bf 0}) - {\bf t}^T {\bf K}^{-1} ({\bf 0}\oplus {\bf t_2})$ (see Supplementary text). As an example, in the bilayer $^2_1$CF family, the $\nu^*_1=\nu^*_2=1$ state is represented by ${\bf K}=\{3,1;1,3\}$ with ${\bf t}^T=(1,1)$ and is observed at $\tilde\nu_{\text{t}}=1/2$ and $\Delta \nu=0$. 

Using the above prescription, we construct the ${\bf K}$-matrices describing some of the observed bilayer states discussed in the main text. For brevity, we use the notation $^a_b(\nu^*_1,\nu^*_2)$ which denotes a bilayer state at CF effective fillings $\nu^*_1$ and $\nu^*_2$ belonging to the $^a_b$CF family. 
Starting with the state $^2_1(1,2)$ at $\tilde\nu_{\text{t}}=7/13$ and $\Delta\nu=-1/13$, the ${\bf K}$-matrix is,
\begin{align}
 ^2_1(1,2)~\text{state}: ~~ {\bf K} = \begin{pmatrix}
        3 & 1 & 1 \\ 1& 3 & 2 \\ 1 & 2 & 3
    \end{pmatrix}, ~{\bf t}=\begin{pmatrix}
        1\\1\\1
    \end{pmatrix}.
\end{align}
Next, for the state $^2_1(2,2)$ at $\tilde\nu_{\text{t}}=4/7$ and $\Delta\nu=0$,
\begin{align}
 ^2_1(2,2)~\text{state}: ~~{\bf K} = \begin{pmatrix}
        3 & 2 & 1 & 1\\ 2&3&1&1 \\ 1&1& 3 & 2 \\ 1&1 & 2 & 3
    \end{pmatrix}, ~{\bf t}=\begin{pmatrix}
        1\\1\\1\\1
    \end{pmatrix}. 
\end{align}
For the state $^2_1(1,3)$ at $\tilde\nu_{\text{t}}=5/9$ and $\Delta\nu=-1/9$, 
\begin{align}
 ^2_1(1,3)~\text{state}: ~~{\bf K} = \begin{pmatrix}
        3 & 1 & 1 & 1 \\ 1& 3 & 2 &2 \\ 1 & 2 & 3 &2 \\ 1&2&2&3
    \end{pmatrix}, ~{\bf t}=\begin{pmatrix}
        1\\1\\1\\1
    \end{pmatrix} .
\end{align}
For the state $^2_1(1,-3)$ at $\tilde\nu_{\text{t}}=2/3$ and $\Delta\nu=-1/3$, 
\begin{align}
 ^2_1(1,-3)~\text{state}: ~~{\bf K} = \begin{pmatrix}
        3 & 1 & 1 & 1 \\ 1& 1 & 2 &2 \\ 1 & 2 & 1 &2 \\ 1&2&2&1
    \end{pmatrix}, ~{\bf t}=\begin{pmatrix}
        1\\1\\1\\1
    \end{pmatrix} .
\end{align}
Finally, the state $^2_1(2,-4)$ at $\tilde\nu_{\text{t}}=2/3$ and $\Delta\nu=-2/9$, 
\begin{align}
 ^2_1(2,-4)~\text{state}: ~~{\bf K} = \begin{pmatrix}
        3 & 2 & 1 & 1 & 1 & 1\\ 2& 3 & 1 & 1& 1& 1 \\ 1&1& 1 & 2 &2 & 2 \\ 1&1 & 2 & 1 &2 &2 \\ 1&1&2&2&1 &2 \\ 1&1&2&2&2&1
    \end{pmatrix}, ~{\bf t}=\begin{pmatrix}
        1\\1\\1\\1\\1\\1
    \end{pmatrix} .
\end{align}
All the properties of the topological order can be extracted from these ${\bf K}$-matrices and their corresponding ${\bf t}$-vectors. In the next section, we use the ${\bf K}$-matrix formalism to find the exciton charges and statistics.
\subsubsection{Exciton charges and statistics}
 It is  possible to find the fundamental quasiparticle excitations, and their statistics for a given ${\bf K}$-matrix. The excitons are then the neutral excitations with quasiparticle charges localized in the two layers. A generic quasiparticle excitation with charge $Q_1$ in layer 1 and $Q_2$ in layer 2 is labeled by the vector ${\bf l}$ with integer components where $\dim {\bf l}=\dim {\bf t_1}+\dim {\bf t_2}$. The quasiparticle charge in each layer is given by (see Supplementary text),
\begin{align}\label{Eq:charge1}
    Q_1 = -e ({\bf t_1}\oplus {\bf 0})^T {\bf K}^{-1} {\bf l},
    \end{align}
   in layer 1 and similarly, in layer 2 it is,
    \begin{align}\label{Eq:charge2}
    Q_2 = -e ({\bf 0}\oplus {\bf t_2})^T {\bf K}^{-1} {\bf l}.
\end{align}
From the above formula, it is evident that a generic excitation charge in the bilayer state, $Q=Q_1+Q_2$, consists of charges localized in both layers. Thus, a fractional charge excitation in the bilayer state is composed of fractional charges localized in the two layers, however, these individual layer charges are not independently created. The exciton charge is computed from the condition $Q_1+Q_2=0$. Additionally, the self statistical phase $\theta_s$ accumulated when a quasiparticle labeled by a vector ${\bf l}$, in the bilayer system is exchanged with the same kind of another quasiparticle, is given by (see Supplementary text),
\begin{align}
    \theta_s = \pi {\bf l}^T{\bf K}^{-1}{\bf l} \mod 2\pi.
\end{align}

In the example considered before, in the bilayer $^2_1$CF family, the $\nu^*_1=\nu^*_2=1$ state allows excitons which are bound states of quasiparticles with charge of $Q_1=-3e/8$ and $Q_2=e/8$ is the two layers, along with qausiholes with charge of $Q_1=-e/8$ and $Q_2=3e/8$. Thus the exciton has a charge of $\pm e/2$, as illustrated in Fig. \ref{fig4}d. We would like to emphasize again that $Q_1$ and $Q_2$ are not to be thought of as independent charges. The $\pm e/2$ exciton acquires a self statistical phase of $\pi$ after exchange, which makes it a fermionic exciton.

An advantage of working in the symmetric representation is that the  charges $Q_1,Q_2$, and their statistics, only depend on the following two integers $\ell_1,\ell_2\in \mathbb{Z}$, 
\begin{align}\label{Eq:qpint}
\ell_1 \equiv \sum_{i=1}^{\dim {\bf t}_1} l_i, ~~~\text{and}~~~  \ell_2 \equiv \sum_{i=1}^{\dim {\bf t}_2} l_{i+\dim {\bf t}_1} .
\end{align}
See Supplementary text for a proof of the above statement. We compute the minimal exciton charge construction corresponding to the ${\bf K}$-matrices given for some of the observed bilayer states discussed in the main text. First, for the state $^2_1(1,2)$ at $\tilde\nu_{\text{t}}=7/13$ and $\Delta\nu=-1/13$, we obtain,
\begin{align}
 \frac{Q_1}{e} = \frac{5}{13} \ell_1 - \frac{1}{13} \ell_2, ~~~\text{and}~~~\frac{Q_2}{e} = \frac{3}{13} \ell_2 - \frac{2}{13}\ell_1,
 \end{align}
 which results in the exciton charge $Q_{\text{exciton}}= \pm e$. This exciton acquires a self statistical phase of $8\pi$ modulo $2\pi$, which makes it a boson. We would like to emphasize that although for $^2_1(1,2)$ state the exciton is bosonic with charge $\pm e$, depairing of such excitons leads to highly non-trivial fractional quasiparticles. Next, for the state $^2_1(2,2)$ at $\tilde\nu_{\text{t}}=4/7$ and $\Delta\nu=0$, the charges are,
 \begin{align}
     \frac{Q_1}{e} = \frac{5}{21} \ell_1 - \frac{2}{21} \ell_2, ~~~\text{and}~~~ \frac{Q_2}{e} = \frac{5}{21} \ell_2 - \frac{2}{21}\ell_1,
    \end{align}
    which results in the exciton charge $Q_{\text{exciton}}= \pm e/3$. In this state, the statistical phase acquired when an exciton goes around another is $\theta_s=4\pi/3$. Next, for the state $^2_1(1,3)$ at $\tilde\nu_{\text{t}}=5/9$ and $\Delta\nu=-1/9$, the charges are,
    \begin{align}
 \frac{Q_1}{e} = \frac{7}{18} \ell_1 - \frac{1}{18} \ell_2, ~~~\text{and}~~~ \frac{Q_2}{e} = \frac{1}{6} \ell_2 - \frac{1}{6}\ell_1,
      \end{align}
      which leads to an exciton charge $Q_{\text{exciton}}= \pm e/2$. The statistical phase for this exciton is $\theta_s=7\pi/2$, or equivalently $3\pi/2$. Although the two states, $^2_1(1,1)$ and $^2_1(1,3)$, share the same exciton charge, they follow different statistics, which highlights the rich internal structure of these fractional excitons. Next, for the state $^2_1(1,-3)$ at $\tilde\nu_{\text{t}}=2/3$ and $\Delta\nu=-1/3$, the charges are,
      \begin{align}
\frac{Q_1}{e} = \frac{5}{12} \ell_1 - \frac{1}{12} \ell_2, ~~~\text{and}~~~ \frac{Q_2}{e} = \frac{1}{4} \ell_2 - \frac{1}{4}\ell_1,
       \end{align}
       which leads to an exciton charge $Q_{\text{exciton}}= \pm e/2$. The statistical phase for this exciton is $\theta_s = 2\pi$ making it a boson. Finally, for the state $^2_1(2,-4)$ at $\tilde\nu_{\text{t}}=2/3$ and $\Delta\nu=-2/9$, the charges are,
       \begin{align}
 \frac{Q_1}{e} = \frac{7}{27} \ell_1 - \frac{2}{27} \ell_2, ~~~\text{and}~~~ \frac{Q_2}{e} = \frac{5}{27} \ell_2 - \frac{4}{27}\ell_1,
\end{align}
which results in the exciton charge $Q_{\text{exciton}}= \pm e/3$. The statistical phase for this exciton is $\theta_s=2\pi$, thus it is bosonic as well. The illustration for some of the exciton constructions is shown in Fig. \ref{fig4}d. As a general rule, for a state belonging to the $^2_1$CF family with effective integer fillings $\nu^*_1$ and $\nu^*_2$, the exciton charge is given by 
\begin{align}
Q_{\text{exciton}}=\pm \frac{e}{\text{gcd}(|\nu^*_1+1|,|\nu^*_2+1|)},    
\end{align}
where the integer valued function gcd$(n_1,n_2)$ computes the greatest common divisor of the integers $n_1$ and $n_2$. Similarly, for integer filled $\nu_1^*,\nu_2^*\in \mathbb{Z}$ states belonging to the $^2_2$CF family, the exciton charge is always $\pm e$.

\subsubsection{Competing states}

At certain special points in the $\nu_{\text{t}}$-$\Delta\nu$ phase space, the observed incompressible states lie on the intersection of the family of curves belonging to multiple $^a_b$CF-type constructions, which makes identifying the underlying structure of the wavefunction ambiguous. Such states, for example at $\nu_{\text{t}}=2/3$ and $\Delta\nu=0$, admit several topologically inequivalent competing orders which, in general, lead to different excitonic charges. Hence, with the aid of future experiments, the ground states can be resolved. 

In this section, we discuss some of these competing states. Starting with the state at $\nu_{\text{t}}=2/3$ and $\Delta\nu=0$, a ${\bf K}$-matrix can be constructed belonging to the $^2_0$CF family,
\begin{align}
    ^2_0(1,1)~\text{state:} ~~{\bf K}=\begin{pmatrix}
        3 & 0 \\ 0 & 3
    \end{pmatrix}, ~{\bf t} =\begin{pmatrix}
        1\\1
    \end{pmatrix}.
\end{align}
An equally possible construction, belonging to the $^2_2$CF family, involves integer filled $\nu^*_1=\nu^*_2=-1$ state. The ${\bf K}$-matrix for such a state is given by,
\begin{align}
    ^2_2(-1,-1)~\text{state:} ~~{\bf K}=\begin{pmatrix}
        1 & 2 \\ 2 & 1
    \end{pmatrix}, ~{\bf t} =\begin{pmatrix}
        1\\1
    \end{pmatrix}.
\end{align}
Naturally, since the two topological orders belong to different equivalence classes, the two descriptions have different physical properties. The exciton charge, for example, in the $^2_0$CF type state is simply a bound state of quasiparticle-quasihole charges in the two layers, which is $Q_{\text{exciton}}=\pm e/3$. The self statistical phase is $\theta_s=2\pi/3$. In the $^2_2$CF type state, however, the charges are given by,
\begin{align}
    \frac{Q_1}{e} = \frac{1}{3} \ell_1 - \frac{2}{3} \ell_2, ~~~\text{and}~~~ \frac{Q_2}{e} = \frac{1}{3} \ell_2 - \frac{2}{3}\ell_1,
\end{align}
which leads to the exciton charge $Q_{\text{exciton}}=\pm e$, with self statistical phase of $\theta_s = 2\pi$, making it a boson. Another set of competing orders exist at $\nu_{\text{t}} = 4/5$ and $\Delta\nu=0$. Belonging to the $^2_0$CF family, the state can be described as,
\begin{align}
    ^2_0(2,2)~\text{state:} ~~{\bf K}=\begin{pmatrix}
        3 & 2 & 0 & 0 \\ 2 & 3 & 0 & 0 \\ 0 & 0& 3 & 2 \\ 0 & 0 & 2 & 3
    \end{pmatrix}, ~{\bf t} =\begin{pmatrix}
        1\\1\\1\\1
    \end{pmatrix}.
\end{align}
Since there is no interlayer flux, the state admits an exciton with charge given simply as $Q_{\text{exciton}}=\pm e/5$. The self statistical phase is $\theta_s=6\pi/5$. At the same time,  the state can  be described 
in terms of the $^2_1$CF family as,
\begin{align}
    ^2_1(-2,-2)~\text{state:} ~~{\bf K}=\begin{pmatrix}
        1 & 2 & 1 & 1 \\ 2 & 1 & 1 & 1 \\ 1 & 1 & 1 & 2 \\ 1 & 1 & 2 & 1
    \end{pmatrix}, ~{\bf t} =\begin{pmatrix}
        1\\1\\1\\1
    \end{pmatrix}.
\end{align}
whose charges in each layer are given by,
\begin{align}
    \frac{Q_1}{e} = \frac{3}{5} \ell_1 - \frac{2}{5} \ell_2, ~~~\text{and}~~~ \frac{Q_2}{e} = \frac{3}{5} \ell_2 - \frac{2}{5}\ell_1,
\end{align}
leading to the exciton charge $Q_{\text{exciton}}=\pm e$, that follows bosonic statistics $\theta_s=2\pi$.

\newpage
\clearpage

\pagebreak
\begin{widetext}
\section{Supplementary Materials}

\begin{center}
\textbf{\large Excitons in the Fractional Quantum Hall Effect}\\
\vspace{10pt}

Naiyuan James Zhang$^{\ast}$,
Ron Q. Nguyen$^{\ast}$, Navketan Batra$^{\ast}$, Xiaoxue Liu$^{\dag}$, 
 Kenji Watanabe, Takashi Taniguchi, D. E. Feldman, and J.I.A. Li$^{\ddag}$

\vspace{10pt}
$^{\ddag}$ Corresponding author. Email: jia$\_$li @brown.edu
\end{center}

\noindent\textbf{This PDF file includes:}
\noindent{Figs. S1 to S4}

\renewcommand{\vec}[1]{\boldsymbol{#1}}

\renewcommand{\thefigure}{S\arabic{figure}}
\def\theequation{S\arabic{equation}}
\def\thetable{S\Roman{table}}
\setcounter{figure}{0}
\setcounter{equation}{0}

\noindent{Supplementary Text}

\noindent{Supplementary Data}

\section{Supplementary Text}

\subsection{Drag current in bilayer Corbino setup}
The philosophy behind the calculation is the same as presented in Section III in Method. Focusing on the FQHE bilayer states belonging to the $^2_1$CF family, we apply an external electric field to layer 1 and compute the backflow current due to interlayer fluxes. As the composite fermions move, they drag along the attached magnetic flux. This in turn creates an effective electric field which results in the drag current.

By convention, we assume the magnetic field to be pointing along the negative $z$ direction. We apply an external electric field $E$ along $x$ which results in the flow of composite fermion current $j_{i,\alpha}$ along $\alpha=x,y$ direction in $i$th layer. Since the composite fermions carry (2+1) fluxes, the induced electric fields due to the motion of fluxes are,
\begin{align}
\delta E_{1,x} = - \frac{2\Phi_0}{ec} j_{1,y} - \frac{\Phi_0}{ec} j_{2,y},~~~~~~~\text{and}~~~~~~~\delta E_{1,y} = \frac{2\Phi_0}{ec} j_{1,x} + \frac{\Phi_0}{ec} j_{2,x},
\end{align}
where $\Phi_0=hc/e$ is the flux quantum. The subscripts of $\delta E_{i,\alpha}$ represent the layer index $i=1,2$ and the direction $\alpha=x,y$. Similarly, the induced electric fields in the second layer are,
\begin{align}
    \delta E_{2,x} = - \frac{\Phi_0}{ec} j_{1,y} - \frac{2\Phi_0}{ec} j_{2,y},~~~~~~~\text{and}~~~~~~~\delta E_{2,y} = \frac{\Phi_0}{ec} j_{1,x} + \frac{2\Phi_0}{ec} j_{2,x}.
\end{align}
Since the external electric field is only applied to the first layer, we compute the current assuming non-zero but small longitudinal conductivity $\sigma^i_{xx}\ne 0$, and quantized transverse conductivity $\sigma_{xy}^{i}=\nu^*_ie^2/h$ for the $i$th layer, where $\nu_i^*\in\mathbb{Z}$ are the effective CF fillings. We have,
\begin{align}
    j_{1,x} &= \sigma^1_{xx} \left[ E - \frac{2\Phi_0}{ec} j_{1,y} - \frac{\Phi_0}{ec} j_{2,y} \right] - \sigma^{1}_{xy} \left[ \frac{2\Phi_0}{ec} j_{1,x} + \frac{\Phi_0}{ec} j_{2,x}  \right], \\
    j_{2,x} &= \sigma^{2}_{xx} \left[ - \frac{\Phi_0}{ec} j_{1,y} - \frac{2\Phi_0}{ec} j_{2,y} \right] - \sigma^{2}_{xy} \left[ \frac{\Phi_0}{ec} j_{1,x} + \frac{2\Phi_0}{ec} j_{2,x} \right],\\
    j_{1,y} &= \sigma^1_{xx} \left[ \frac{2\Phi_0}{ec} j_{1,x} + \frac{\Phi_0}{ec} j_{2,x} \right] + \sigma^{1}_{xy} \left[ E - \frac{2\Phi_0}{ec} j_{1,y} - \frac{\Phi_0}{ec} j_{2,y} \right],\\
j_{2,y} &= \sigma^{2}_{xx} \left[ \frac{\Phi_0}{ec} j_{1,x} + \frac{2\Phi_0}{ec} j_{2,x} \right] + \sigma^{2}_{xy} \left[ - \frac{\Phi_0}{ec} j_{1,y} - \frac{2\Phi_0}{ec} j_{2,y} \right].
\end{align}

The coupled equations can be solved for the currents $j_{i,\alpha}$ in each layer and each direction up to the leading order in $\sigma_{xx}$. From Ohm's law, $j_{x}\sim \sigma_{xx}E$ in each layer, thus up to leading order, we obtain,
\begin{align}\label{Eq:drag_supplemantal}
    j_{1,y} = \frac{\nu^*_1(1+2\nu^*_2)}{1+2(\nu^*_1+\nu^*_2)+3\nu^*_1\nu^*_2} \frac{e^2}{h} E,~~~~~~~\text{and}~~~~~~~j_{2,y} = -\frac{\nu^*_1\nu^*_2}{1+2(\nu^*_1+\nu^*_2)+3\nu^*_1\nu^*_2} \frac{e^2}{h} E.
\end{align}
Assuming $\sigma^1_{xx}\approx \sigma^2_{xx}\equiv \sigma_{xx}$ we solve for the current densities along $x$ direction,
\begin{align}
    j_{1,x} = \frac{\nu_1^{*2}+(1+2\nu^*_2)^2}{\left[1+2(\nu^*_1+\nu^*_2)+3\nu^*_1\nu^*_2\right]^2} \sigma_{xx} E,~~~~~~~\text{and}~~~~~~~j_{2,x} = - \frac{\nu^*_1+\nu^*_2+2(\nu_1^{*2}+\nu_2^{*2})}{\left[1+2(\nu^*_1+\nu^*_2)+3\nu^*_1\nu^*_2\right]^2} \sigma_{xx}  E.
\end{align}
The condition of perfect drag is now
\begin{equation}
    \nu_1^{*2}+(1+2\nu_2^*)^2=\nu_1^*+\nu_2^*+2(\nu_1^{*2}+\nu_2^{*2}).
\end{equation}
If the drive layer is layer 2 then the perfect drag condition becomes
\begin{equation}
    \nu_2^{*2}+(1+2\nu_1^*)^2=\nu_1^*+\nu_2^*+2(\nu_1^{*2}+\nu_2^{*2}).
\end{equation}
The sum of the two equations yields
\begin{equation}
    (1+\nu_1^*)^2+(1+\nu_2^*)^2=0,
\end{equation}
that is, $\nu_1^*=\nu_2^*=-1$. Beyond this limit, the mechanism of this section cannot produce perfect drag for a drive in each of the two layers.

\subsection{Filling fractions, quasiparticle charges, and exciton statistics from K-matrix formalism}
In this subsection, we give the derivation of the expressions for total filling $\tilde\nu_{\text{t}}$ and the difference of fillings $\Delta \nu$ stated in the main text. We also derive the formulae for quasiparticle charges localized in each layer and exciton statistics. For a symmetric ${\bf K}$-matrix constructed via the prescription presented in the Methods section, and their corresponding ${\bf t_1}$ and ${\bf t_2}$ vectors, the action for the bilayer system, involving $\dim {\bf t_1} + \dim {\bf t_2}$ emergent $U(1)$ gauge fields $\{a^i_{\mu}\}$, is given as,
\begin{align}\label{Eq:action}
    S = \frac{e^2}{\hbar c} \epsilon^{\lambda\mu\nu} \int d^2x dt \left[ \frac{1}{4\pi} {\bf a}_{\lambda}^T {\bf K} \partial_{\mu}{\bf a}_{\nu} -\frac{A^1_{\lambda}}{2\pi} ({\bf t_1}\oplus {\bf 0})^T \partial_{\mu}{\bf a}_{\nu} - \frac{A^2_{\lambda}}{2\pi} ({\bf 0}\oplus {\bf t_2})^T \partial_{\mu}{\bf a}_{\nu}  \right], 
\end{align}
where the operation $\oplus$ denotes direct sum, and we represent the emergent gauge fields as a vector ${\bf a}^T_{\mu}\equiv (a^1_{\mu},a^2_{\mu},\dots)$.  The current density in the two layers, $J_1^{\lambda}=(e^2/h c)\epsilon^{\lambda\mu\nu}({\bf t_1}\oplus{\bf 0})^T\partial_{\mu}{\bf a}_{\nu}$ and $J_2^{\lambda}=(e^2/hc)\epsilon^{\lambda\mu\nu}({\bf 0}\oplus{\bf t_2})^T\partial_{\mu}{\bf a}_{\nu}$, couple to vector potentials $A^1_{\mu}$ and $A^2_{\mu}$ respectively which result from an externally applied displacement field, therefore, $A^i_{\mu}=A^0_{\mu}+\delta A^i_{\mu}$ for layers $i=1,2$. By varying the action, one finds the equation of motion, corresponding to each gauge field, that is compactly written in matrix form,
\begin{align}
    A^1_{\mu}{\bf t_1} \oplus A^2_{\mu}{\bf t_2} = {\bf K}{\bf a}_{\mu} ~~~\Rightarrow~~~ {\bf K}^{-1} \left( A^1_{\mu}{\bf t_1} \oplus A^2_{\mu}{\bf t_2}\right) = {\bf a}_{\mu}.
\end{align}
We now integrate out all the emergent $U(1)$ gauge fields $\{a^i_{\mu}\}$, and using the symmetric property of the ${\bf K}$-matrix, one obtains an effective action,
\begin{align}
    S_{\text{eff}}=-\frac{e^2}{4\pi \hbar c} \epsilon^{\lambda\mu\nu} \int d^2x dt &\big[ A^1_{\lambda} \partial_{\mu}A^1_{\nu} ({\bf t_1}\oplus {\bf 0})^T{\bf K}^{-1}({\bf t_1}\oplus {\bf 0}) + A^2_{\lambda} \partial_{\mu}A^2_{\nu} ({\bf 0}\oplus {\bf t_2})^T{\bf K}^{-1}({\bf 0}\oplus {\bf t_2}) \nonumber \\
    &+ A^1_{\lambda}\partial_{\mu}A^2_{\nu} ({\bf t_1}\oplus{\bf 0})^T{\bf K}^{-1}({\bf 0}\oplus{\bf t_2}) + A^2_{\lambda}\partial_{\mu}A^1_{\nu} ({\bf 0}\oplus{\bf t_2})^T{\bf K}^{-1}({\bf t_1}\oplus{\bf 0}) \big].
\end{align}
Since the two layers are coupled to different vector potentials, leading order variation of the externally applied displacement field gives the filling of the individual layers $\tilde\nu_1$ and $\tilde\nu_2$,
\begin{align}
    S_{\text{eff}}= -\frac{e^2}{4\pi \hbar c} \epsilon^{\lambda\mu\nu} \int d^2x dt &\big[ A^0_{\lambda}\partial_{\mu}A^0_{\nu} ({\bf t_1}\oplus{\bf t_2})^T{\bf K}^{-1}({\bf t_1}\oplus{\bf t_2}) + \left( A^0_{\lambda}\partial_{\mu}\delta A^1_{\nu} + \delta A^1_{\lambda}\partial_{\mu} A^0_{\nu} \right) ({\bf t_1}\oplus{\bf t_2})^T{\bf K}^{-1}({\bf t_1}\oplus{\bf 0}) \nonumber \\
    &+ \left( A^0_{\lambda}\partial_{\mu}\delta A^2_{\nu} + \delta A^2_{\lambda}\partial_{\mu} A^0_{\nu} \right) ({\bf t_1}\oplus{\bf t_2})^T{\bf K}^{-1}({\bf 0}\oplus{\bf t_2}) + \mathcal{O}(\delta A)^2 \big].
\end{align}
From here we obtain the response of the bilayer state with respect to the change in the displacement field. Consequently, the individual layer filling fractions are given by,
\begin{align}
    \tilde\nu_1 = ({\bf t_1}\oplus{\bf t_2})^T {\bf K}^{-1} ({\bf t_1}\oplus {\bf 0})~~~~~\text{and}~~~~~~\tilde\nu_2 = ({\bf t_1}\oplus{\bf t_2})^T {\bf K}^{-1} ({\bf 0}\oplus {\bf t_2}).
\end{align}
The total filling and the difference of fillings are then defined as $\tilde \nu_{\text{t}}\equiv\tilde\nu_1+\tilde\nu_2$ and $\Delta\nu\equiv \tilde\nu_1-\tilde\nu_2$, which lead to the expressions used in the main text. 

The same setup can also be used to extract the quasiparticle charges in each layer by including the source terms. We define $\dim {\bf t}_1 + \dim {\bf t}_2$ current densities $\{j^i_{\mu}\}$ associated to the emergent gauge fields $\{a^i_{\mu}\}$, each of which couple to the emergent gauge field, thus we include
\begin{align}
    S_{\text{source}} = -\int d^2x dt ~\delta^{\mu\nu}{\bf j}_{\mu}^T{\bf a}_{\nu},
\end{align}
the source term to the action Eq. \ref{Eq:action}. Here we represent the sources as a vector ${\bf j}_{\mu}^T\equiv (j^1_{\mu},j^2_{\mu},\dots)$. The equation of motion for the system with the source term is now given as,
\begin{align}
    {\bf j}^{\lambda} = \frac{e^2}{hc} \epsilon^{\lambda\mu\nu} \left[ -\partial_{\mu}(A^1_{\nu}{\bf t}_1\oplus A^2_{\nu}{\bf t}_2) + {\bf K}\partial_{\mu}{\bf a}_{\nu} \right] ~~~~\Rightarrow~~~~{\bf K}^{-1}{\bf j}^{\lambda} = \frac{e^2}{hc} \epsilon^{\lambda\mu\nu} \left[ -{\bf K}^{-1}\partial_{\mu}(A^1_{\nu}{\bf t}_1\oplus A^2_{\nu}{\bf t}_2) + \partial_{\mu}{\bf a}_{\nu} \right] 
\end{align}
Using these equations of motion, the actual current densities, $J^{\mu}_1$ and $J^{\mu}_2$ in the two layers, are then computed in terms of the sources. Quasiparticles are created by inserting sources, ${\bf j}^0({\bf r})=-e{\bf l}\delta({\bf r}-{\bf r}_0)$, labelled by a vector ${\bf l}$ of dimensions $\dim {\bf l}=\dim {\bf t}_1+\dim{\bf t}_2$ and integer components. Since the charge associated with the emergent gauge field is different from the charge in the original EM field, we obtain the charges,
\begin{align}
    Q_1 = -e ({\bf t}_1\oplus {\bf 0})^T{\bf K}^{-1}{\bf l}, ~~~~~~\text{and}~~~~~~Q_2 = -e ({\bf 0}\oplus {\bf t}_2)^T{\bf K}^{-1}{\bf l},
\end{align}
localized in the two layers. A generic excitation in the bilayer state, labeled by an integer vector ${\bf l}$, consists of charges localized in both layers. The exciton condition is given by $Q_1 + Q_2 = 0$, i.e., for an exciton labeled by ${\bf l}_e$, the condition translates to $({\bf t}_1\oplus{\bf t}_2)^T{\bf K}^{-1}{\bf l}_e=0$. Strictly speaking, the exciton condition is also accompanied by a constraint that ${\bf l}_e$ correspond to a minimal $Q_{1,2}\ne 0$, since more complicated states may have neutral excitations that are not excitons.

Following similar arguments as presented above, one finds the self-statistical phase $\theta_s$ accumulated when one exciton, labeled by ${\bf l}_e$, in the bilayer system, is exchanged with the same kind of another exciton,
\begin{align}
\theta_s = \pi {\bf l}_e^T {\bf K}^{-1} {\bf l}_e \mod 2\pi.
\end{align} 
We notice that the exciton condition $({\bf t}_1\oplus{\bf t}_2)^T{\bf K}^{-1}{\bf l}_e=0$ may have distinct solutions corresponding to integer valued ${\bf l}_e$. All such minimal excitons, corresponding to the distinct ${\bf l}_e$, however, lead to the same statistical phase modulo $2\pi$. We show this in the next section.

\subsection{Some useful properties of the bilayer ${\bf K}$-matrices}

In this section, we exploit some of the useful properties of the bilayer ${\bf K}$ matrix in the symmetric representation. We use the prescription given in Section III in Method. As a reminder, the ${\bf K}$-matrices, and their corresponding ${\bf t}$-vectors, for the bilayer states are described in terms of blocks in Eq. \ref{Eq:Kmat}. The blocks ${\bf K}_1$ and ${\bf K}_2$, along with their respective charge vectors ${\bf t}_1$ and ${\bf t}_2$, describe the CF filled $\nu_1^*,\nu_2^*\in\mathbb{Z}$ monolayer states respectively. Now, in the symmetric representation, the charge vectors take the form, ${\bf t}^T=(1,\dots,1)$, depending on their respective dimensions, i.e., $\dim {\bf t}_1$ and $\dim {\bf t}_2$ respectively. In this representation, the monolayer ${\bf K}$-matrices take the form,
\begin{align}
    {\bf K}_i = \sigma_i {\bf I} + p_i {\bf J}, ~~~~\text{with}~~~~\sigma_i=\pm 1,~p_i\in \mathbb{Z},  
\end{align}
where the matrices ${\bf I}$ and ${\bf J}$ are the identity and constant matrix of ones respectively. The subscript $i=1,2$ denotes the layer index. Naturally, depending on the layer index, $i=1,2$ the monolayer matrices have a dimension of $\dim{\bf t}_i\times \dim {\bf t}_i$. The simple form of the monolayer ${\bf K}$-matrices admits a simple form of their inverses,
\begin{align}
    {\bf K}_i^{-1} = \sigma_i {\bf I} - \frac{\sigma_ip_i}{p_i \dim {\bf K}_i + \sigma_i} {\bf J}.
\end{align}
Using these results, the structure of the inverse of the bilayer ${\bf K}$-matrix, constructed via the prescription in Eq. (\ref{Eq:Kmat}), is given as,
\begin{align}
    {\bf K}^{-1} = \begin{pmatrix}
        \sigma_1 {\bf I} + \lambda_1 {\bf J}_1 & \lambda' {\bf J} \\ 
        \lambda' {\bf J}^T & \sigma_2 {\bf I} + \lambda_2 {\bf J}_2
    \end{pmatrix},~~~~~~~\text{with}~~~~~~~~\sigma_i = \pm 1,\text{ and }\lambda_i,\lambda' \in \mathbb{Q}. 
\end{align}

The structure of the inverse of the bilayer ${\bf K}$-matrix allows us to make certain claims about the dependence of quasiparticle charge and statistics on the choice of ${\bf l}$-vector. The quasiparticle charge in each layer is given by Eqs. (\ref{Eq:charge1}), and (\ref{Eq:charge2}). \textit{Claim 1:} In the symmetric representation, the quasiparticle charges in each layer $Q_1$ and $Q_2$ depend only on the two integers $\ell_1,\ell_2\in\mathbb{Z}$, defined in Eq. (\ref{Eq:qpint}). \textit{Proof:} To prove, we use the structure of the inverse of the ${\bf K}$-matrix and write,
\begin{align}
    Q_1 &= -e ({\bf t}_1\oplus {0})^T {\bf K}^{-1} {\bf l} =  -e \sum_{i,j=1}^{n} K_{ij}^{-1} l_j  -e \sum_{i=1}^{n}\sum_{j=1+n}^{n+m} K_{ij}^{-1} l_j \\
    &= -e \sum_{i,j=1}^{n} \left( \sigma_1 \delta_{ij} + \lambda_1  \right) l_j  -e n \lambda' \sum_{j=1}^{m} l_j = -e \left[ \left( \sigma_1 + n \lambda_1\right)  \ell_1  + n \lambda' \ell_2 \right],
\end{align}
where for convinience we defined $n\equiv \dim{\bf t}_1$ and $m\equiv \dim{\bf t}_2$. Since the ${\bf K}$-matrix is symmetric, and the two diagonal blocks have the same form, the argument for $Q_2$ is the same. $\blacksquare$

\textit{Claim 2:} In the symmetric representation, for the quasiparticle represented by the vector ${\bf l}$, the self statistics $\theta_s$ depend only on the two integers $\ell_1,\ell_2\in\mathbb{Z}$. \textit{Proof:} Again, the structure of the inverse of ${\bf K}$-matrix allows one to prove the claim. Using symmetric property, we have,
\begin{align}
     {\bf l}^T {\bf K}^{-1} {\bf l} &= \sum_{i,j=1}^{n} l_il_j K_{ij}^{-1} + \sum_{i,j=1+n}^{m+n} l_il_j K_{ij}^{-1} + 2\sum_{i=1}^{n}\sum_{j=1+n}^{m+n} l_il_j K_{ij}^{-1} \\
     & = \sum_{i,j=1}^{n} l_il_j\left( \sigma_1 \delta_{ij} + \lambda_1 \right) + \sum_{i,j=1+n}^{m+n} l_il_j \left( \sigma_2 \delta_{ij} + \lambda_2 \right) + 2 \lambda'  \sum_{i=1}^{n} l_i \sum_{j=1+n}^{m+n} l_j \\
     & = \sigma_1 \sum_{i=1}^{n} l_i^2 + \sigma_2 \sum_{i=1}^{m} l_{i+n}^2 + \lambda_1 \ell_1^2  + \lambda_2 \ell_2^2 + 2 \lambda'  \ell_1 \ell_2.
\end{align}
We need to show that the first two terms in the above expression do not change $\theta_s = \pi {\bf l}^T {\bf K}^{-1} {\bf l}$ modulo $2\pi$, when the components of the vector ${\bf l}$ change such that they result in the same $Q_1$ and $Q_2$. From our previous proof, we showed that the charges only depend on the integers $\ell_1$ and $\ell_2$ rather than all the components of ${\bf l}$. Therefore, we show that $\theta_s$ remains invariant as the vector ${\bf l}$ changes keeping $\ell_{1,2}$ fixed. These transformations are achieved by repeatedly replacing $l_k \rightarrow l_k + s_1$ and $l_j \rightarrow l_j-s_1$ for some $1\leq k,j\leq n$ and similarly, replacing $l_k \rightarrow l_k + s_2$ and $l_j \rightarrow l_j-s_2$ for some $n+1\leq k,j\leq m+n$ where $s_1,s_2\in \mathbb{Z}$. These transformations result in,
\begin{align}
    \sigma_1 \sum_{i=1}^{n} l_i^2 ~~(\text{mod } 2) &~\longrightarrow ~  \sigma_1 (l_k+s_1)^2 + \sigma_1 (l_j-s_1)^2 +\sigma_1 \sum_{i=1,i\ne j,k}^{n} l_i^2  ~~(\text{mod } 2) \\
    &=  2\sigma_1 s_1 (l_k+l_j) + 2\sigma_1 s_1^2 + \sigma_1 \sum_{i=1}^{n} l_i^2 ~~(\text{mod } 2)  = \sigma_1 \sum_{i=1}^{n} l_i^2 ~~(\text{mod } 2),
\end{align}
where in the last equality, we make use of the fact that $\sigma_1 = \pm 1$ and $s_1,l_i\in \mathbb{Z}$. Similarly, under these changes the expression $\sigma_2 \sum_{i=1}^{m} l_{i+n}^2$ modulo $2$ is invariant therefore, $\theta_s = \pi {\bf l}^T {\bf K}^{-1} {\bf l}$ modulo $2\pi$ is determined by the integers $\ell_1$ and $\ell_2$. $\blacksquare$

A similar argument can be used to compare mutual statistics of two excitons with identical $\ell_1$, $\ell_2$ but different ${\bf l}$-vectors. We discover that the statistical angle is uniquely defined up to $\pi$. This phase difference has no physical consequences since an exchange of non-identical objects does not return the system to the initial quantum state. If, on the other hand, one exciton makes a full circle around another exciton then the statistical-phase difference doubles $\pi \rightarrow 2\pi$ and is irrelevant.

\begin{figure*}
\includegraphics[width=1\linewidth]{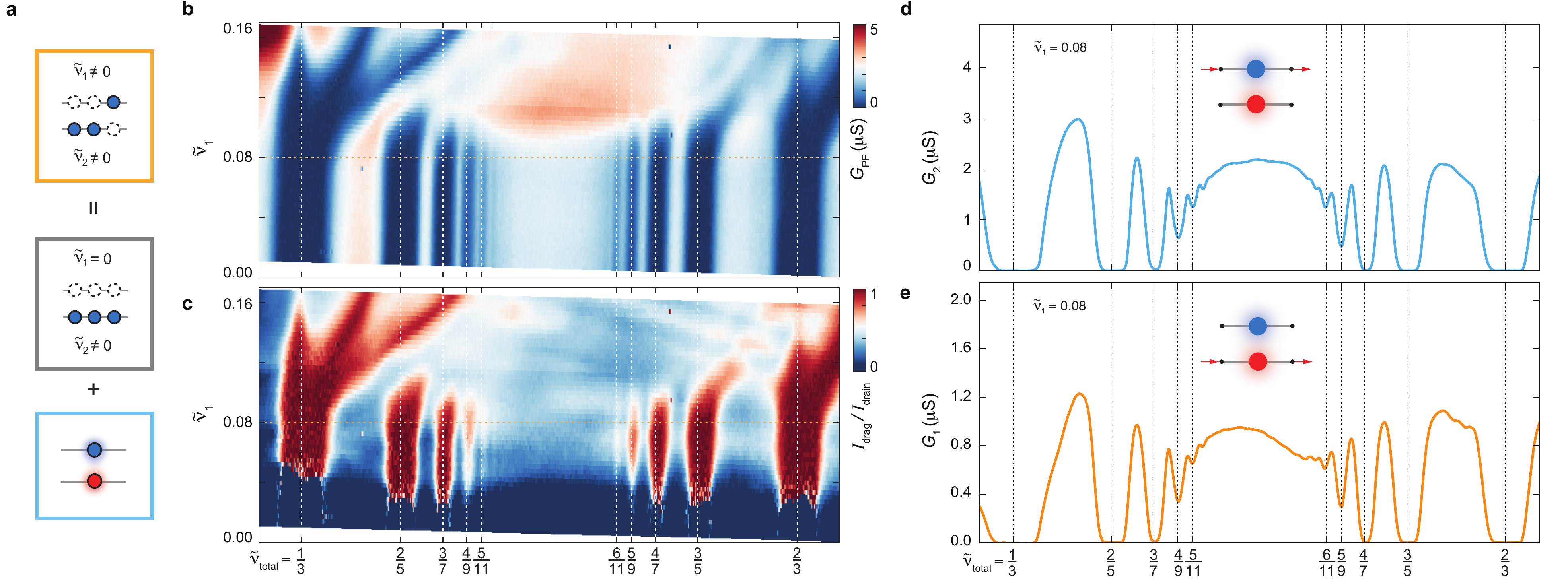}
\caption{\label{2componentJain}{\bf{2-component FQHE effect with \CFii\ excitons. }} (a) Schematic diagram of the \CFii\ exciton construction. (b) $G_{\text{PF}}$ and (c) drag ratio $I_{\text{drag}}/I_{\text{drive}}$ measured as a function of $\tilde{\nu}_{\text{total}}$ and $\tilde{\nu}_1$. The orange horizontal line indicates the same orange line in Fig.~\ref{fig3}b. (d-e) Bulk conductance measured across (d) layer 1 and (e) layer 2 as a function of $\tilde{\nu}_{\text{total}}$ at $\tilde{\nu}_1 = 0.08$. Varying $\tilde{\nu}_{\text{total}}$ at $\tilde{\nu}_1 = 0.08$ defines the orange line in panel b and c, which corresponds to the same trajectory as the orange line in Fig.~\ref{fig3}b. Along the orange line, LL filling in layer 1 remains a constant, $\tilde{\nu}_1 =0.08$. As such, it is remarkable that bulk conductance measured across layer 1, $G_1$, exhibits the same oscillation as that of layer 2 $G_2$. Since $\tilde{\nu}_1$ is a constant, only $\tilde{\nu}_2$ changes along the orange line. The fact that $G_1$ varies between highly conductive and insulating with varying $\tilde{\nu}_2$ provides unambiguous evidence for interlayer excitonic pairing. Moreover, layer 1 is conductive with non-zero $G_1$ away from \CFii\ excitonic orders. This is an independent confirmation that layer 1 is populated with free charge carriers and the quantum Hall bilayer is outside the single-layer regime.  }
\end{figure*}

\begin{figure*}
\includegraphics[width=1\linewidth]{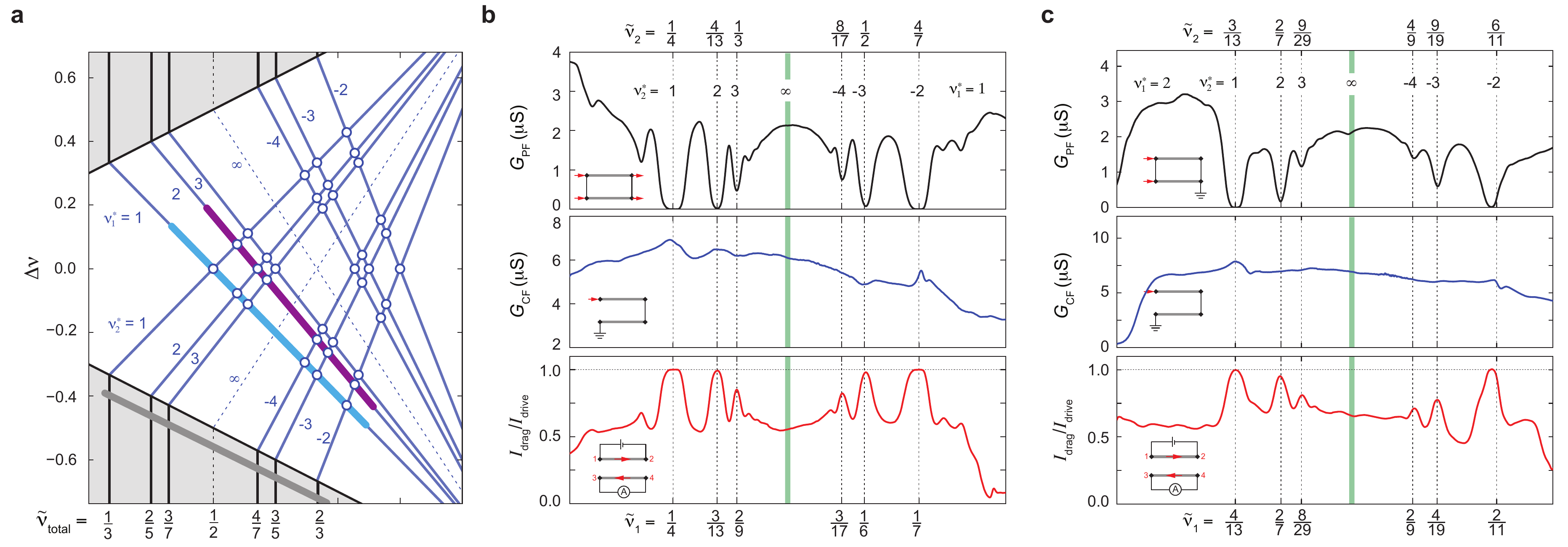}
\caption{\label{CF21}{\bf{\CFi\ sequence of FQHE. }} (a) Schematic diagram of \CFi\ states across a portion of \nutot-\dnu\ map. Blue solid lines denote integer values of $^2_1\nu_{i}^{\ast}$. Light blue line marks $^2_1\nu_{1}^{\ast} = 1$, whereas purple line denotes $^2_1\nu_{2}^{\ast}=2$. (b-c) $G_{\text{PF}}$ (top), $G_{\text{CF}}$ (middle) and drag ratio (bottom) measured along the (b) light blue line and (c) purple line in panel (a).  }
\end{figure*}

\begin{figure*}
\includegraphics[width=1\linewidth]{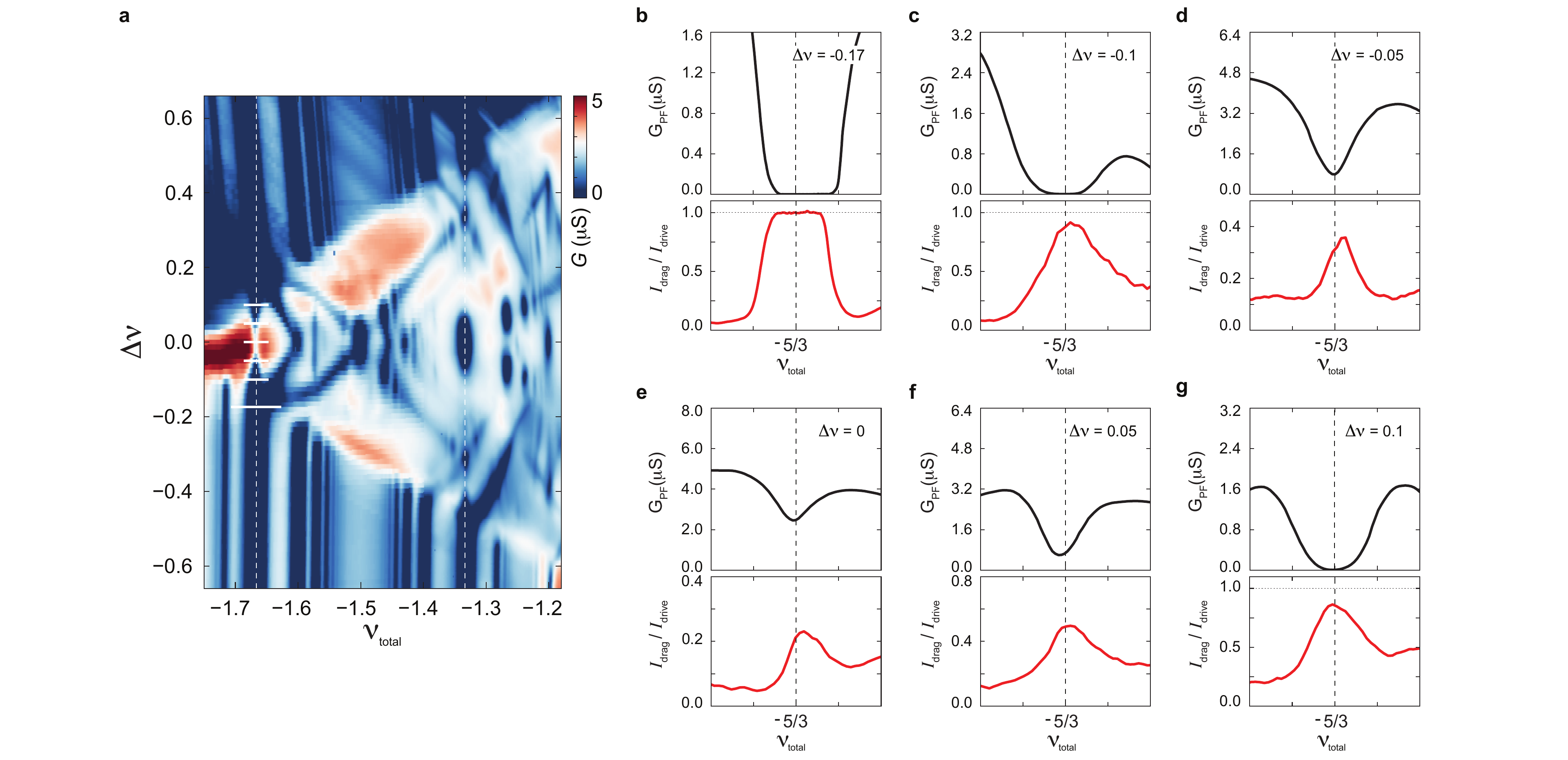}
\caption{\label{1over3}{\bf{\dnu-evolution at $\tilde{\nu}_{\text{total}}= 1/3$. }}  (a) $G_{\text{PF}}$ versus $\nu_{\text{total}}$ and $\Delta\nu$. (b-g) $G_{\text{PF}}$ (top) and $I_{\text{drag}}/I_{\text{drive}}$ (bottom) versus $\nu_{\text{total}}$ near $\nu_{\text{total}}=-5/3$ at different $\Delta\nu$: (b) $-0.17$, (c) $-0.1$, (d) $-0.05$, (e) $0$, (f) $0.05$ (g) $0.1$ (marked as white solid lines in panel a). All measurements are performed at $B=12$ T, $T=20$ mK. Near \dnu $=0$, the stability of the 2-component FQHE $\tilde{\nu}_{\text{total}}= 1/3$ is partially suppressed, which is evidenced by the non-zero $G_{\text{PF}}$ and drag ratio that deviates from one. Nevertheless, $G_{\text{PF}}$ always exhibits a local minimum, and drag ratio always displays a local maximum $\tilde{\nu}_{\text{total}}= 1/3$ regardless of \dnu. We argue that the ground state order is described by the (333) wavefunction throughout the \dnu\ range. The non-zero $G_{\text{PF}}$ near \dnu$=0$ arises from unpaired vortices at the base temperature of the dilution fridge, suggesting that the transition temperature of the exciton condensate is suppressed near \dnu $=0$. Such suppression is analogous to the observed behavior of the exciton condensate at total filling of one ~\cite{Zeng2023solid,Li2017superfluid,Champagne.08b}.  }
\end{figure*}

\begin{figure*}
\includegraphics[width=0.8\linewidth]{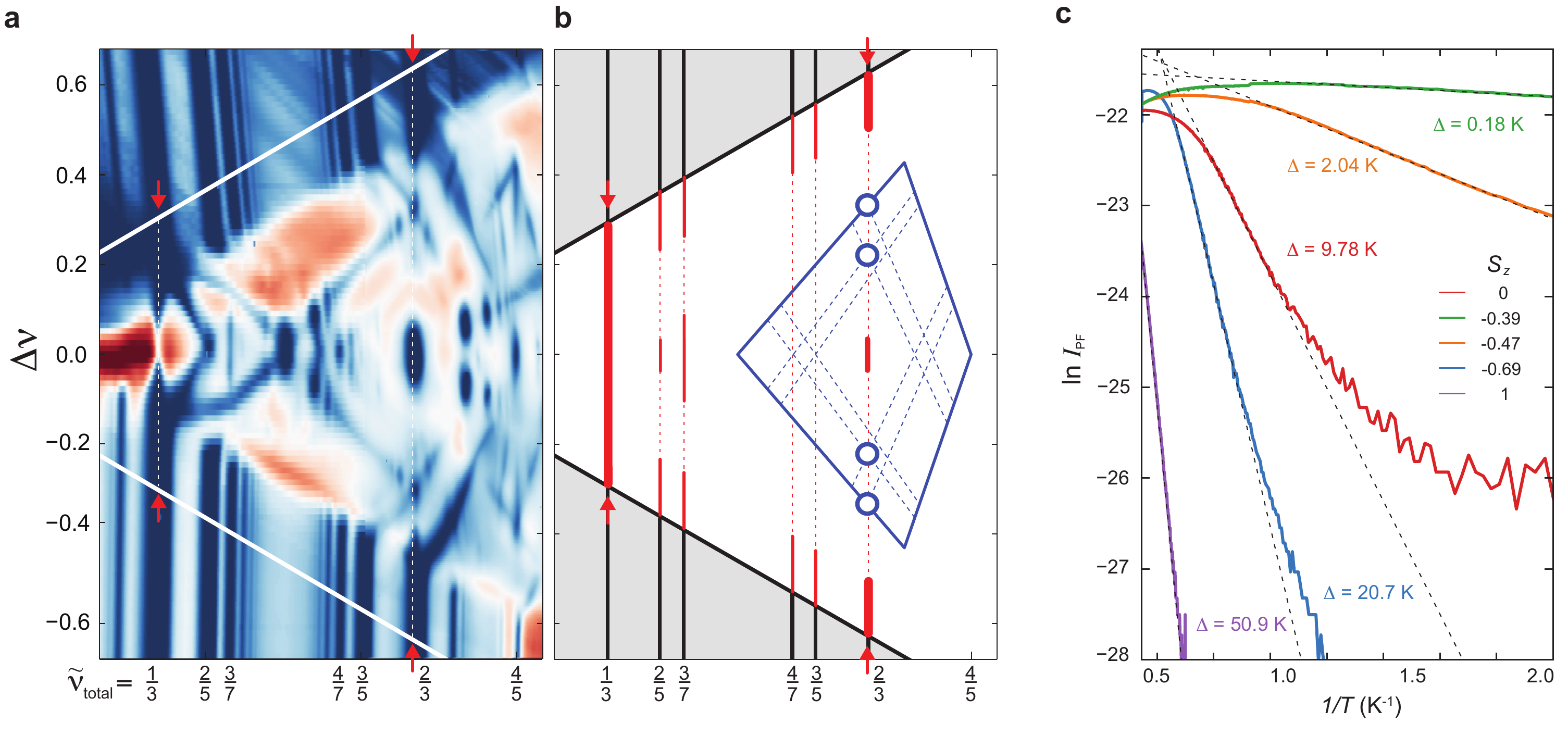}
\caption{\label{2over3}{\bf{Charge gaps and neutral modes at $\tilde{\nu}_{\text{total}}=2/3$. }} (a) $G_{\text{PF}}$ as a function of \dnu\ and $\tilde{\nu}_{\text{total}}$. Vertical dashed lines and red arrows highlight two interesting filling fractions, $\tilde{\nu}_{\text{total}} = 1/3$ and 2/3. (b) Schematic diagram marks the location of \CFii\ and \CFi\ states with red and blue lines, respectively. Open circles mark different FQHE states appearing at $\tilde{\nu}_{\text{total}} = 2/3$.  (c) Arrhenius plot of PF conductance measured at $\tilde{\nu}_{\text{total}} = 2/3$ and different \dnu.  }
\end{figure*}

\end{widetext}

\end{document}